\documentclass[a4paper,11pt]{article}
\pdfoutput=1 

\usepackage{jheppub} 

\usepackage[T1]{fontenc} 

\title{\boldmath Precision measurement of the neutrino velocity with the ICARUS detector in the CNGS beam}

\author[a]{The ICARUS Collaboration\\M.~Antonello}
\author[b]{B.~Baiboussinov}
\author[c]{P.~Benetti}
\author[c]{F.~Boffelli}
\author[c]{E.~Calligarich}
\author[a]{N.~Canci}
\author[b]{S.~Centro}
\author[e]{A.~Cesana}
\author[f]{K.~Cie\'slik}
\author[g]{D.~B.~Cline}
\author[d]{A.~G.~Cocco}
\author[f]{A.~Dabrowska}
\author[b]{D.~Dequal}
\author[h]{A.~Dermenev}
\author[c]{R.~Dolfini}
\author[b]{C.~Farnese}
\author[b]{A.~Fava}
\author[i]{A.~Ferrari}
\author[d]{G.~Fiorillo}
\author[b]{D.~Gibin}
\author[h]{S.~Gninenko}
\author[b]{A.~Guglielmi}
\author[f]{M.~Haranczyk}
\author[n]{J.~Holeczek}
\author[h]{A.~Ivashkin}
\author[n]{J.~Kisiel}
\author[n]{I.~Kochanek}
\author[m]{J.~Lagoda}
\author[n]{S.~Mania}
\author[c]{A.~Menegolli}
\author[b]{G.~Meng}
\author[c]{C.~Montanari}
\author[g]{S.~Otwinowski}
\author[c]{A. Piazzoli}
\author[o]{P.~Picchi}
\author[c]{F.~Pietropaolo,\note{Corresponding author: francesco.pietropaolo@pd.infn.it}}
\author[p]{P.~Plonski}
\author[c]{A.~Rappoldi}
\author[c]{G.~L.~Raselli}
\author[c]{M.~Rossella}
\author[a,i]{C.~Rubbia}
\author[e]{P.~Sala}
\author[l]{E.~Scantamburlo}
\author[e]{A.~Scaramelli}
\author[a]{E.~Segreto}
\author[q]{F.~Sergiampietri}
\author[a]{D.~Stefan}
\author[m]{J.~Stepaniak}
\author[a,m]{R.~Sulej}
\author[f]{M.~Szarska}
\author[e]{M.~Terrani}
\author[b]{F.~Varanini}
\author[b]{S.~Ventura}
\author[a]{C.~Vignoli}
\author[g]{H.G.~Wang}
\author[g]{X.~Yang}
\author[f]{A.~Zalewska}
\author[c]{A.~Zani}
\author[p]{K. Zaremba}
\author[i]{\\and\\P.~Alvarez~Sanchez}
\author[r]{L.~Biagi} 
\author[r]{R.~Barzaghi}
\author[r]{B.~Betti}
\author[s]{L.-G.~Bernier}
\author[t]{G.~Cerretto} 
\author[r]{C.~De~Gaetani}
\author[u]{H.~Esteban}
\author[v]{T.~Feldmann,\note{Present address: TimeTech, Stuttgart, Germany}}
\author[i]{J.~D.~Gonzalez~Cobas}
\author[r]{D.~Passoni} 
\author[t]{V.~Pettiti}
\author[r]{L.~Pinto}
\author[i]{J.~Serrano}
\author[a]{P.~Spinnato} 
\author[r]{M.~G.~Visconti} 
\author[i]{T.~Wlostowski}

\affiliation[a]{INFN - Laboratori Nazionali del Gran Sasso, Assergi (AQ), Italy}
\affiliation[b]{Dipartimento di Fisica e Astronomia e INFN, Universit\`a di Padova, Via Marzolo 8, I-35131,Padova, Italy}
\affiliation[c]{Dipartimento di Fisica  e INFN, Universit\`a di Pavia, Via Bassi 6, I-27100, Pavia, Italy}
\affiliation[d]{Dipartimento di Scienze Fisiche e INFN, Universit\`a Federico II, Napoli, Italy}
\affiliation[e]{INFN, Sezione di Milano e Politecnico, Via Celoria 16, I-20133, Milano, Italy}
\affiliation[f]{H.Niewodnicza\'nski Institute of Nuclear Physics, Krak\'ow, Poland}
\affiliation[g]{Department of Physics and Astronomy, University of California, Los Angeles, USA}
\affiliation[h]{Institute for Nuclear Research of the Russian Academy of Sciences, prospekt 60-letiya Oktyabrya 7a, Moscow 117312, Russia}
\affiliation[i]{CERN, European Laboratory for Particle Physics, CH-1211 Geneve 23, Switzerland}
\affiliation[l]{Universit\`a di L'Aquila, via Vetoio, Localit\`a Coppito, I-67100 L'Aquila, Italy}
\affiliation[m]{National Centre for Nuclear Research, A. Soltana 7, 05 400 Otwock/Swierk, Poland}
\affiliation[n]{IInstitute of Physics, University of Silesia, Uniwersytecka 4, 40-007 Katowice, Poland}
\affiliation[o]{INFN Laboratori Nazionali di Frascati, Via Fermi 40, I-00044, Frascati, Italy}
\affiliation[p]{Institute of Radioelectronics, Warsaw Univ. of Technology, Nowowiejska 15/19, 00 665 Warsaw, Poland}
\affiliation[q]{Dipartimento di Fisica e INFN, Universit\`a di Pisa, Largo Bruno Pontecorvo 3, I-56127, Pisa, Italy}
\affiliation[r]{DIIAR-Politecnico di Milano, Piazza Leonardo da Vinci 32, I-20133 Milano, Italy}
\affiliation[s]{METAS Federal Office of Metrology, Lindenweg 50, Bern-Wabern, Switzerland}
\affiliation[t]{Optics Division, INRIM (Istituto Nazionale di Ricerca Metrologica), Torino, Italy}
\affiliation[u]{Time Department, Real Instituto y Observatorio de la Armada (ROA), San Fernando, Spain}
\affiliation[v]{Physikalisch-Technische Bundesanstalt (PTB), Bundesallee 100, D-38116 Braunschweig, Germany}

\abstract{
During May 2012, the CERN-CNGS neutrino beam has been operated for two weeks for a total of $\sim1.8 \times 10^{17}$ p.o.t., with the proton beam made of bunches, few ns wide and separated by 100 ns. This  beam structure allows a very accurate time of flight measurement of neutrinos from CERN to LNGS on an event-by-event basis.

Both the ICARUS-T600 PMT-DAQ and the CERN-LNGS timing synchronization have been substantially improved for this campaign, taking advantage of additional independent GPS receivers, both at CERN and LNGS as well as of the deployment of the ``White Rabbit'' protocol both at CERN and LNGS. 

The ICARUS-T600 detector has collected 25 beam-associated events; the corresponding time of flight has been accurately evaluated, using all different time synchronization paths.

The measured neutrino time of flight is compatible with the arrival of all events with speed equivalent to the one of light: the difference between the expected value based on the speed of light and the measured value is $\delta t  = tof_c - tof_\nu = 0.10 \pm 0.67_{stat.} \pm 2.39_{syst.}$ ns. This result is in agreement with the value previously reported by the ICARUS Collaboration, $\delta t  = 0.3 \pm 4.9_{stat.}  \pm  9.0_{syst.}$  ns, but with improved statistical and systematic accuracy.}

\begin{document} 
\maketitle
\flushbottom

\section{Indroduction}
\label{sec:intro}
Neutrino time of flight measurements were triggered by the OPERA Collaboration announcement~\cite{[01]} of the surprising result indicating a superluminal propagation speed of CNGS neutrinos from CERN to LNGS. 

Cohen and Glashow~\cite{[02]} argued that such superluminal neutrinos should lose energy by producing photons and $e^+ e^-$ pairs, through $Z_0$ mediated processes, analogous to Cherenkov radiation. The ICARUS Collaboration reported a negative result in the search for superluminal Cherenkov-like pairs inside its large LAr-TPC detector~\cite{[03]} exposed to the CNGS beam. No candidate event was found, setting a tight negative limit on $(v_\nu - c)/c$ of $2.5 \times 10^{-8}$ at 90 \% C.L.
Moreover the ICARUS Collaboration performed a first time of flight measurement of neutrinos from CERN to LNGS with a 3 ns wide bunched beam shortly operated at the end of 2011 CNGS run. The result, $\delta t  = tof_c - tof_\nu = 0.3 \pm 4.9_{stat.}  \pm  9.0_{syst.}$ ns~\cite{[04]}  was compatible with the simultaneous arrival of all events with speed equal to that of light, in a striking difference with the reported initial result of OPERA~\cite{[01]}.

At the end of the 2011 CNGS bunched beam campaign, CERN and the LNGS experiments (Borexino, ICARUS, LVD, OPERA) agreed to perform a new dedicated campaign with the aim of improving the neutrino time of flight measurement significance. This new measurement was performed from May 10$^{th}$ to May 24$^{th}$ 2012 with the CERN-SPS accelerator operated in a dedicated bunched mode with $\sim 10^{12}$ p.o.t (protons on target) per spill~\cite{[05]}. 
The neutrino time of flight measurement consists in recording any neutrino induced interaction time at LNGS and relating it to the transit time of a proton bunch at a BCT monitor along the CNGS TT40 transfer tunnel at CERN, occurring $\sim 2.44$ ms earlier. Precise geodetic estimation of the CERN-LNGS distance allows calculating the actual neutrino velocity.

The ICARUS result on 2011 data has been confirmed with improved precision by the other LNGS experiments~\cite{[06],[26]} in the 2012 CNGS bunched beam campaign. We hereby report the experimental measurements of the neutrino velocity with the ICARUS detector, obtained combining the accurate determination of the distance and time of flight with the direct observation of either neutrino events inside the detector or neutrino associated muons from the surrounding rock.  Improved high-accuracy GPS measurements have provided the determination of the distance from CERN to LNGS and up to the ICARUS detector with an uncertainty of centimetres; an upgraded GPS-based timing system at CERN and LNGS allowed a time of flight measurement with systematic uncertainties of a few nanoseconds.

\section{The CNGS neutrino bunched beam}
\label{sec:cngs}

\begin{figure}[htbp]
\centering 
\includegraphics[width=.90\textwidth]{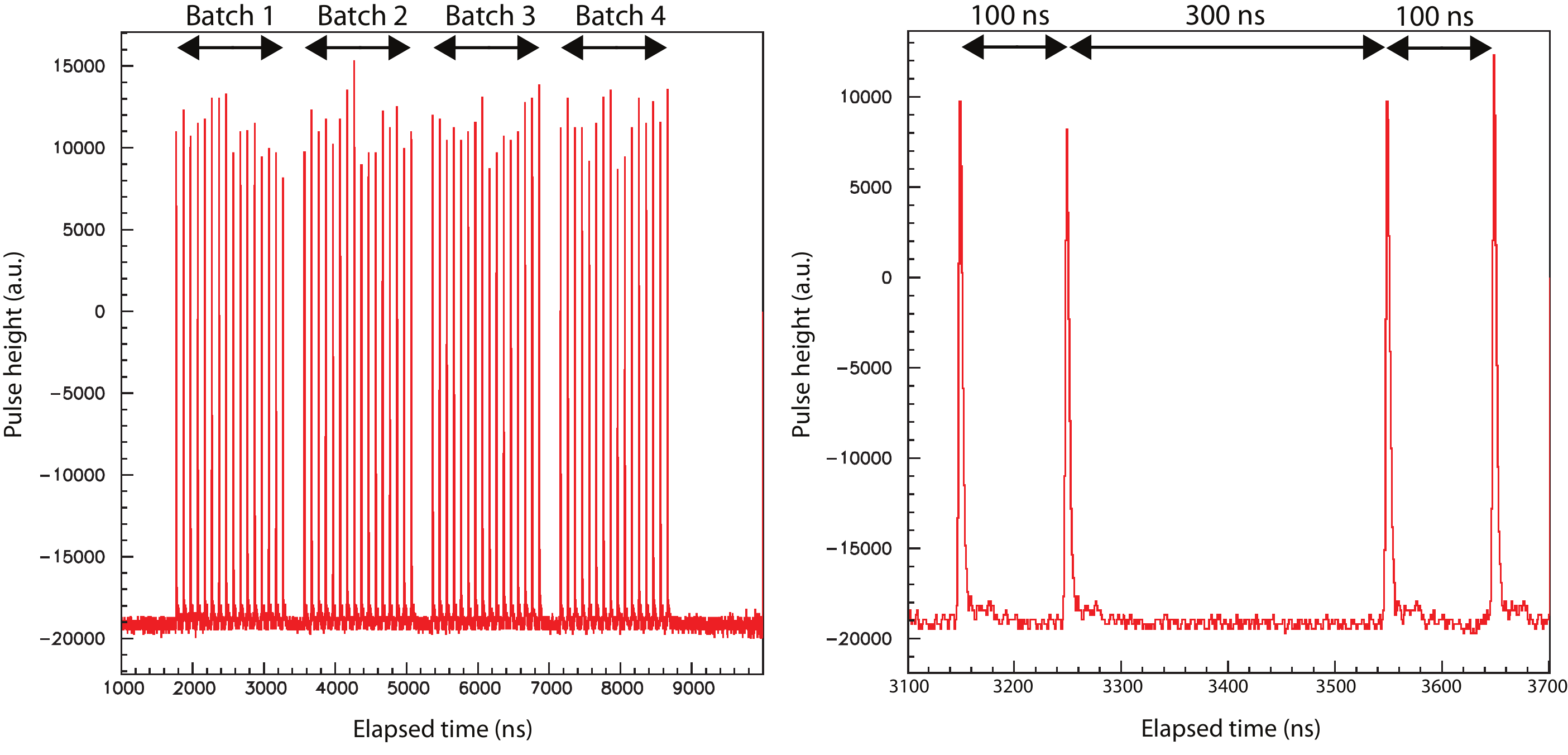}
\caption{\label{fig:bunches} CNGS proton beam structure dedicated to neutrino time of flight measurement. Typical width of each proton pulse is $\sim 4$ ns (FWHM).}
\end{figure}

The CNGS proton beam structure for the 2012 neutrino time of flight run is shown in Figure~\ref{fig:bunches}. It was based on LHC-like proton extractions, with a single extraction per SPS super-cycle (13.2 s), 4 batches per extraction separated by 300 ns, and 16 proton bunches per batch separated by 100 ns; each bunch had a narrow width of $\sim 4$ ns FWHM (1.8 ns rms). This very tightly bunched beam structure allowed a very accurate neutrino time of flight measurement on an event-by-event basis. A total of $\sim 1.8 \times 10^{17}$ p.o.t. were delivered to CNGS allowing the collection of several tens of neutrino induced events, with energy in the 10-40 GeV range, by the experiments located at LNGS at 731 km distance.

\section{Synchronization between CERN and LNGS}
\label{sec:synchro}

\begin{figure}[htbp]
\centering 
\includegraphics[width=.90\textwidth]{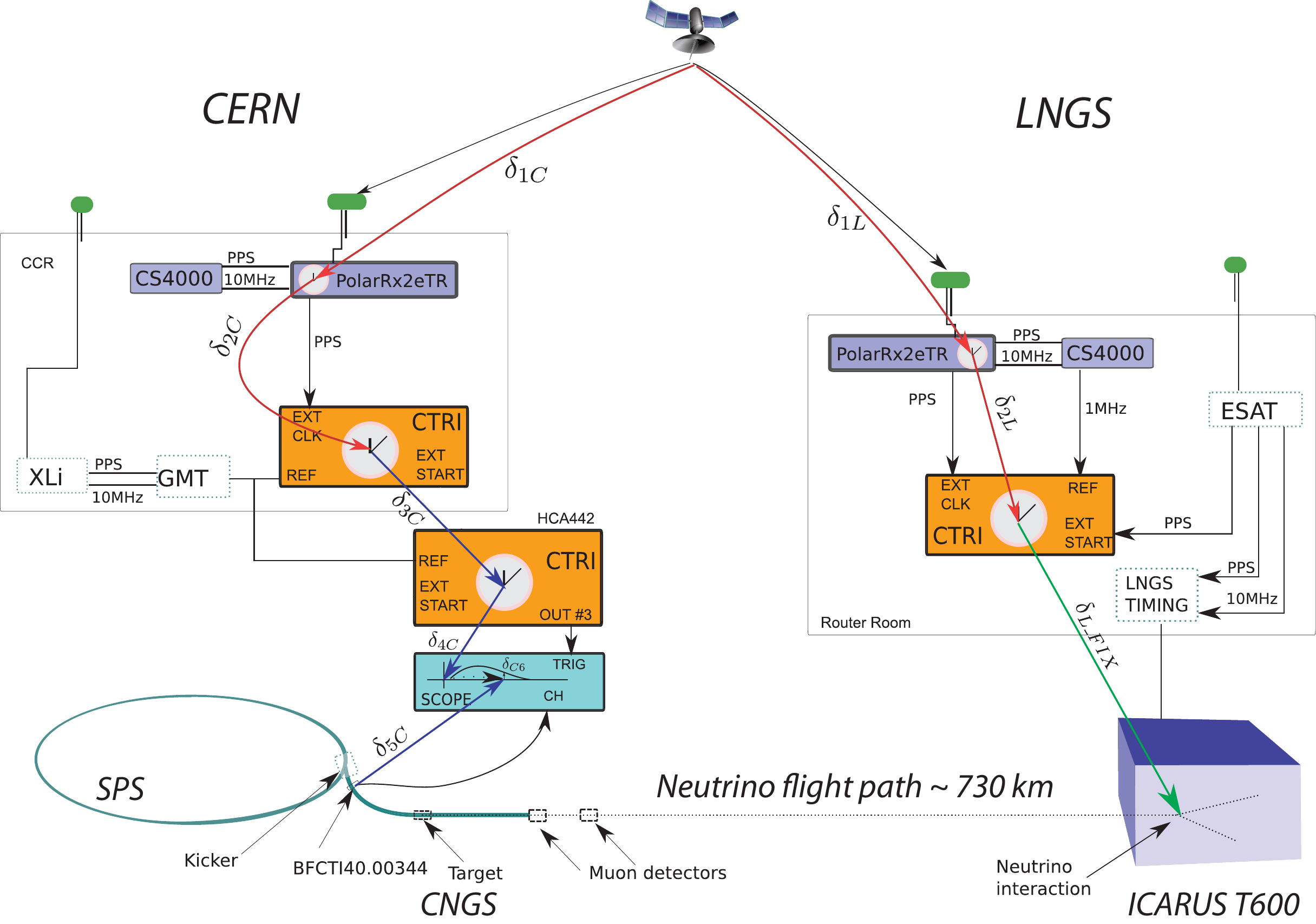}
\caption{\label{fig:timelink} Schematics of the 2011 CERN-LNGS timing synchronization. The origin of the neutrino time
of flight measurement is at BFCTT40. For ICARUS, the arrival reference point is the T600 entry wall.}
\end{figure}

A detailed description of the CERN and LNGS timing systems and their synchronizations, prepared for the 2011 campaign and used also in 2012, is given in Ref.~\cite{[01],[07]}. A schematic picture of the timing system layout, including all delays is shown in Figure~\ref{fig:timelink}. The origin of the neutrino velocity measurement is referred to the Beam Current Transformer (BCT) detector (BFCTI.400344), located $743.391 \pm  0.002$ m upstream of the CNGS neutrino target.

The proton beam time-structure at the BCT is recorded by a 1 GS/s Wave Form Digitiser (WFD) Acqiris DP110~\cite{[08]}, triggered by the SPS kicker magnet signal. At every extraction, the BCTFI.400344 waveform  is stored into the CNGS database. Every acquisition is time-tagged with respect to the SPS timing system, associating each neutrino event at LNGS to a precise proton bunch.

The absolute UTC timing signal at LNGS is provided every second (PPS) by a GPS system ESAT 2000 disciplining  a Rubidium oscillator~\cite{[09],[10]}, operating on the surface Laboratory. A copy of this signal is sent underground every ms (PPmS) and used in ICARUS to provide the absolute time-stamp to the recorded events.

In order to improve the $\sim 100$ ns time accuracy of the standard GPS at CERN and LNGS, the OPERA collaboration and CERN have installed, both at CERN and LNGS, two new identical systems composed of a GPS receiver for time-transfer applications Septentrio PolarRx2e operating in ``common-view'' mode and a Cs atomic clock Symmetricom Cs4000~\cite{[01]}. The Cs4000 oscillator provides the reference frequency to the PolarRx2e receiver, and a CTRI device logs every second the difference in time between the 1PPS outputs of the standard GPS receiver, that drives the SPS timing system (XLi in Figure~\ref{fig:timelink}) and of the more precise PolarRx2e, with 0.1 ns resolution. The two systems were calibrated, resulting in a time base difference of 2.3 $\pm$ 0.9 ns between the CERN and LNGS PolarRx2e receivers. The stability of this calibration was recently re-examined leading to an increased associated error of $\pm$ 2.0 ns~\cite{[07]}. 

The timing signal (PPmS), distributed by the LNGS laboratory, consists of a TTL positive edge (3 ns rise time) sent out every ms and followed, after 200 $\mu s$, by the coding of the absolute time related to the leading edge. This signal is generated in the external laboratory and sent to the underground hall B via $\sim$8 km fibre optics. This introduces a delay, periodically calibrated following a double path procedure very similar to the one devised by the OPERA experiment for their first calibration in 2006~\cite{[01]}.

\subsection{Additional synchronization between CERN and LNGS}
\begin{figure}[htbp]
\centering 
\includegraphics[width=.75\textwidth]{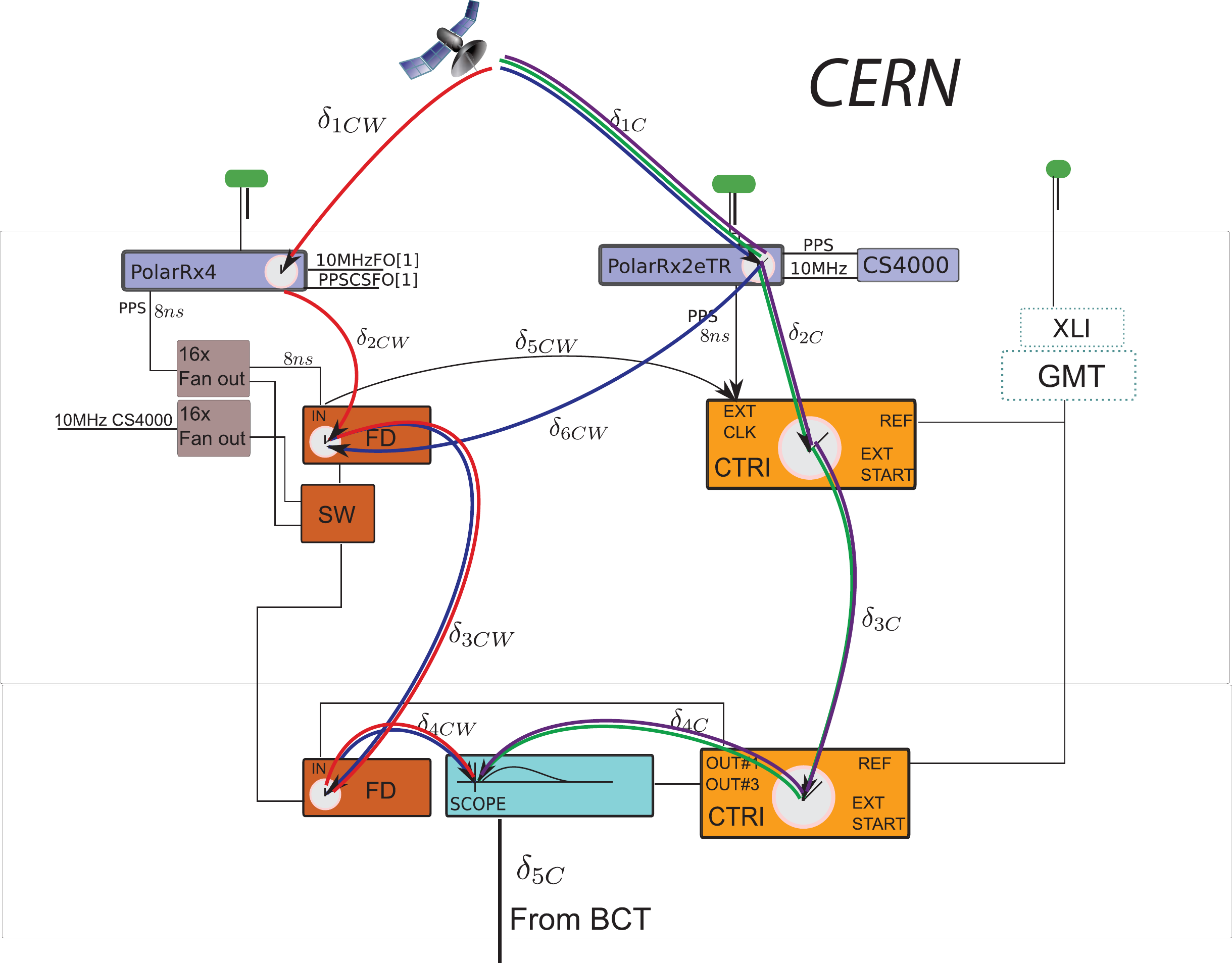}
\includegraphics[width=.85\textwidth]{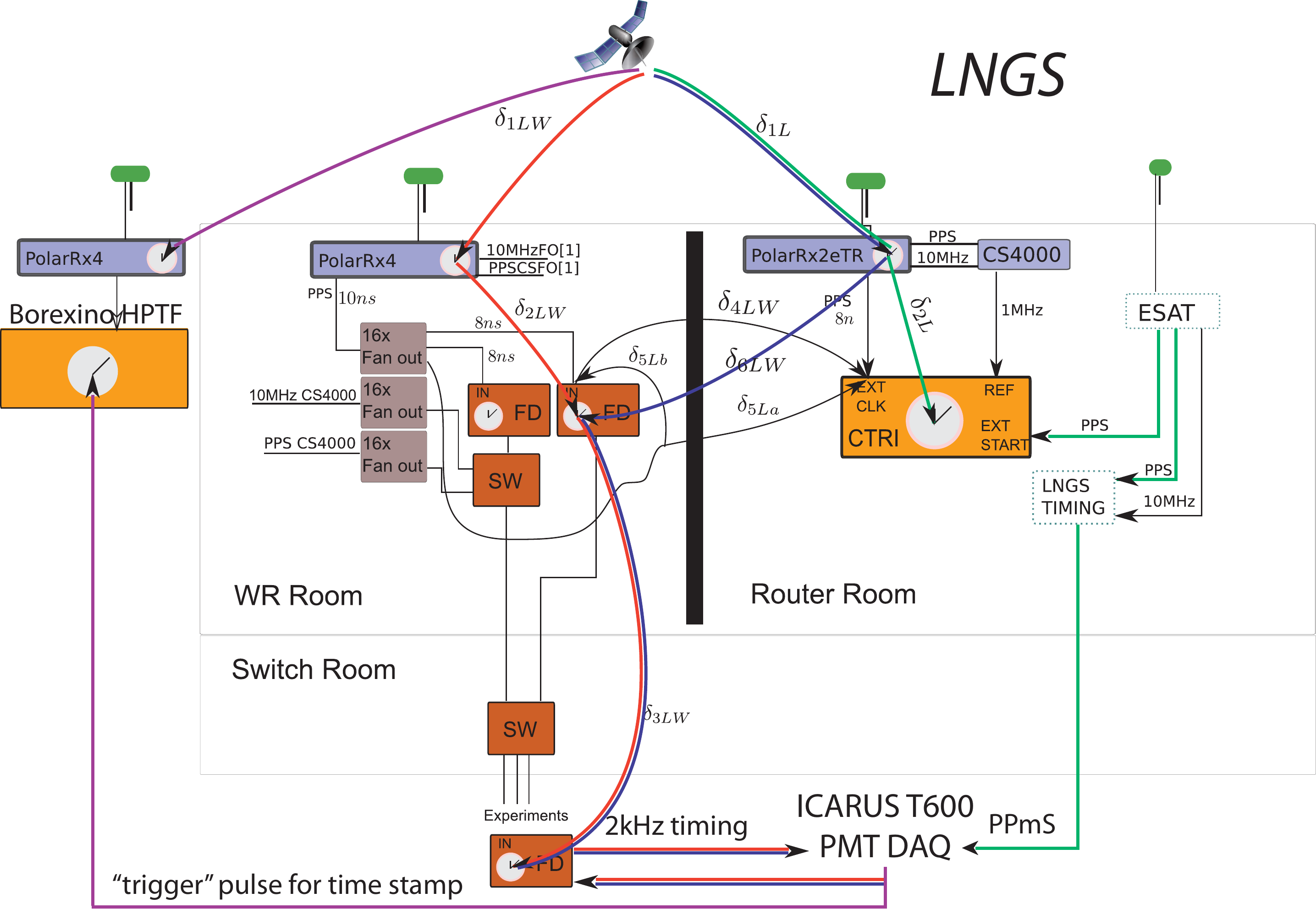}
\caption{\label{fig:layout} Schematics of the 2012 CERN-LNGS time synchronization, including the 2011 classical path
(green), the WR system taking advantage of both the new PolarRx4 (red) and old PolarRx2e (blue), and the
HPTF Borexino facility (violet). Top: layout at CERN. Bottom: layout at LNGS.}
\end{figure}

For the 2012 CNGS bunched beam run, additional CERN-LNGS synchronization systems have been set-up. The new overall layout at CERN and LNGS is depicted in Figure~\ref{fig:layout} top and bottom, respectively.

Under the CERN responsibility, with the aim of providing redundancy and inter-calibration, two new additional Septentrio PolarRx4 GPS receivers, optimized for time-transfer applications, have been installed at CERN and LNGS. As the PolarRx2e receivers, they operate in ``common-view'' mode and are connected to the same Cs atomic clocks Symmetricom Cs4000 used to provide the reference frequency to the PolaRx2e receivers. The inter-calibration of the new PolarRx4 receivers is estimated to be stable within $\pm$ 2.0 ns~\cite{[07]}.

As shown in Figure~\ref{fig:layout}, both the new PolarRx4 and the old PolarRx2e synchronization paths are connected to the DAQ systems at CERN and LNGS, through the ``Classic'' 2011 protocol. In addition to it, a new independent system for timing distribution was  deployed both at CERN and at LNGS. It is based on a recent, still under development, open source protocol, called ``White Rabbit'' (WR)~\cite{[11]}, whose main purpose is to constantly and accurately monitor the propagation delay of any signal along the optical path, connecting all Nodes (PC's provided with WR hardware interface and running the WR protocol firmware) of the WR system.
The WR system is expected to intrinsically correct for any change of the propagation chain delay, thus avoiding the need of periodic calibrations of the optical fibres described in the previous section. Any Node participating in the WR system is thus phase-locked with all the others, with accuracy and stability much better than one ns. The WR protocol allows the distribution of timing signals at various frequencies (e.g. PPS, PPmS) as well as the time-stamping of any pulse generated by the DAQ systems connected to WR Nodes. 

In addition to the above set-up, the Borexino collaboration implemented at LNGS a new timing system (High Precision Timing Facility, HPTF)~\cite{[12]} based on an independent additional PolarRx4 GPS receiver. This installation, located at the LNGS outside laboratory, provides the time stamping of signals propagated from the experiments along the optical fibre-bundle connecting the underground and the external Labs. For the synchronization with CERN, the HPTF relies on the PolarRx2 GPS receiver installed at CERN; the related inter-calibration was performed by INRiM and ROA~\cite{[12]} Institutes. Thanks to a common agreement among Borexino, LVD and ICARUS, the ICARUS experiment had access to the HPTF facility, contributing to the installation of this system with dedicated signal converters and time stamping modules.

\section{Neutrino time of flight measurement with ICARUS}
\label{sec:nutof}
The ICARUS T600 detector consists of a large cryostat split into two identical, adjacent modules with internal dimensions $3.6 \times 3.9 \times 19.6$ m$^3$ filled with about 760 tons of ultra-pure liquid Argon~\cite{[13],[14]}. Each module houses two TPCs separated by a common cathode.  A uniform electric field (E$_{drift}$ = 500 V/cm) is applied. There are three parallel planes of wires, 3 mm apart with lengths up to 9 m, facing the drift volume 1.5 m long. By appropriate voltage biasing, the first two planes provide signals in a non-destructive way allowing to collect the ionization charge on the third plane. There are in total 53248 channels.  Wires are oriented on each plane at different angles $(0^\circ, \pm  60^\circ)$ with respect to the horizontal direction. Combining the wire coordinate on each plane at a given drift time, a three-dimensional image of the ionizing event is reconstructed. A remarkable resolution of $\sim 1$ mm$^3$ is uniformly achieved over the whole detector active volume ($\sim 340$ m$^3$ corresponding to 476 t).

Scintillation light in LAr is abundantly produced by ionizing events ($\sim 2.5 \times 10^4$ photons/MeV at 128 nm wave length); it exhibits  two distinct decay components: the fast component has a decay time of 6 ns and amounts for $\sim 25 \%$ of the total light emission. In the ICARUS LAr-TPC the scintillation is recorded with 74 photomultipliers (PMT: ETL 9357FLA) of 8 inches diameter, organized in horizontal arrays of 9 PMTs each, located behind the wire chambers. In the West module, one array is mounted behind each wire chamber (1L, 1R) at a height of 1.58 m; in the East module, there are three arrays for each chamber (2L$_{upper}$, 2L$_{middle}$, 2L$_{lower}$, 2R$_{upper}$, 2R$_{middle}$, 2R$_{lower}$) at a height of 0.60, 1.58 and 2.56 m respectively. The PMT spacing in each array is 2 m. All PMTs are deposited with wavelength shifter (tetra-phenyl-butadiene, TPB) to detect the 128 nm VUV light. The overall estimated quantum efficiency is about 4 $\%$.

The sums of the signals from the PMT arrays are used for the ICARUS global trigger and to locate the event within the drift volume (``T=0''). The trigger threshold, set at about 100 photoelectrons, allows full detection efficiency for events with energy deposition as low as few hundreds MeV. CNGS neutrinos are recorded requiring a coincidence between a 60 $\mu s$ gate, opened according to the ``Early Warning Signal''~\cite{[01]} for proton extraction from CERN-SPS, and the PMT-Sum signals. Given the geometry of the ICARUS LAr-TPC and the PMT spacing, in the case of multi-GeV CNGS neutrino induced events, several hundred photoelectrons are produced in the PMT's closer to the ionizing event, within 1 ns from the ionization process. This feature makes the ICARUS LAr-TPC very well suited for timing measurement. In addition, the possibility to visually scan the associated 3D event image permits to measure the path of the photons from the interaction vertex to the PMT location with mm accuracy (i.e. sub-ns accuracy on photon propagation time).

\subsection{The ICARUS PMT DAQ for neutrino time of flight measurement}
\begin{figure}[htbp]
\centering 
\includegraphics[width=0.90\textwidth]{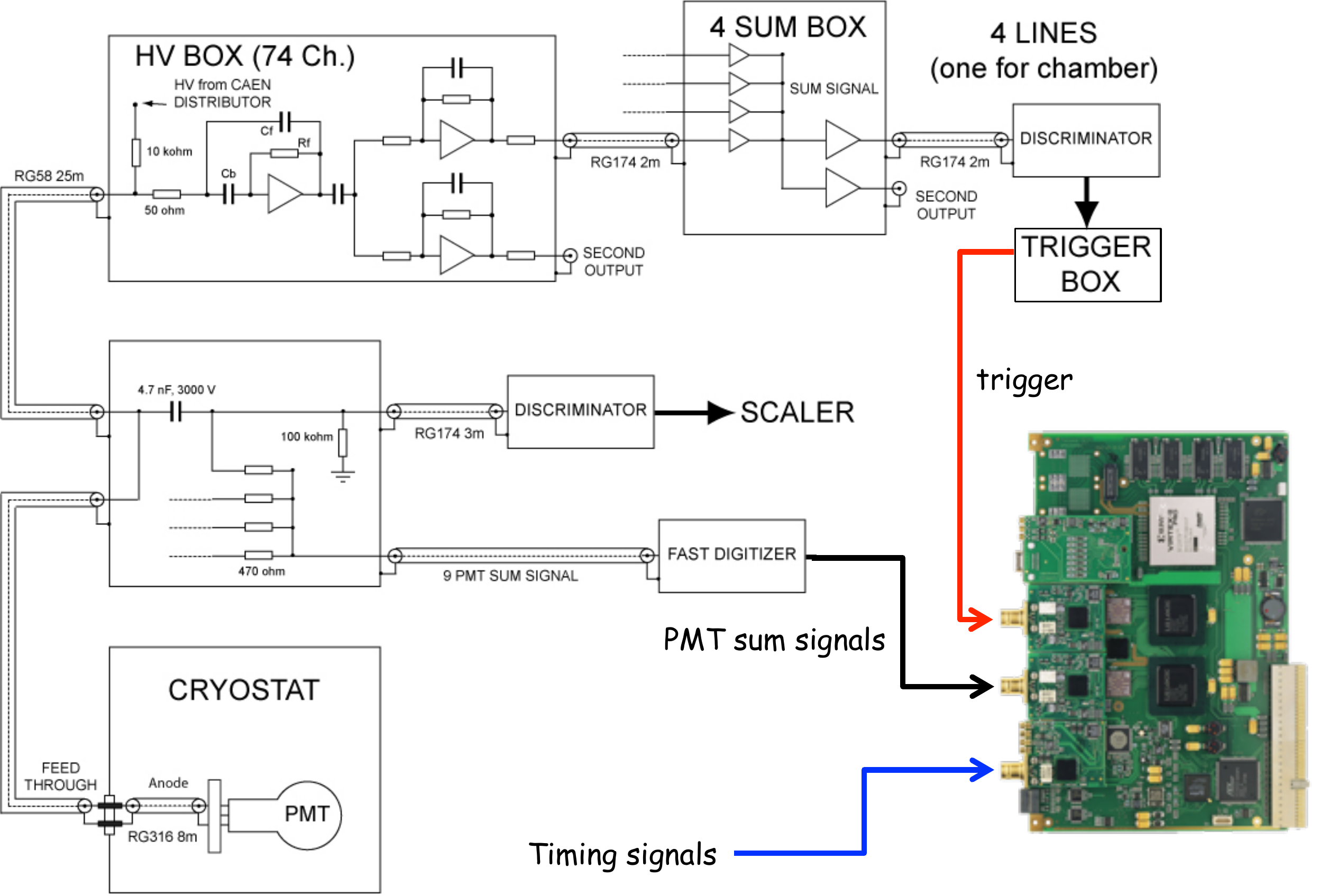}
\caption{\label{fig:t600pmt} Schematics of the 2012 ICARUS PMT-DAQ specifically designed for neutrino time of flight
measurement. The latter is replicated through a dedicated fan-out setup and in order to be sent to the HPTF and to WR facilities for time stamping. The jitter between replicas has been certified to be of the order of 0.1 ns. }
\end{figure}

\begin{figure}[htbp]
\centering 
\includegraphics[width=0.75\textwidth]{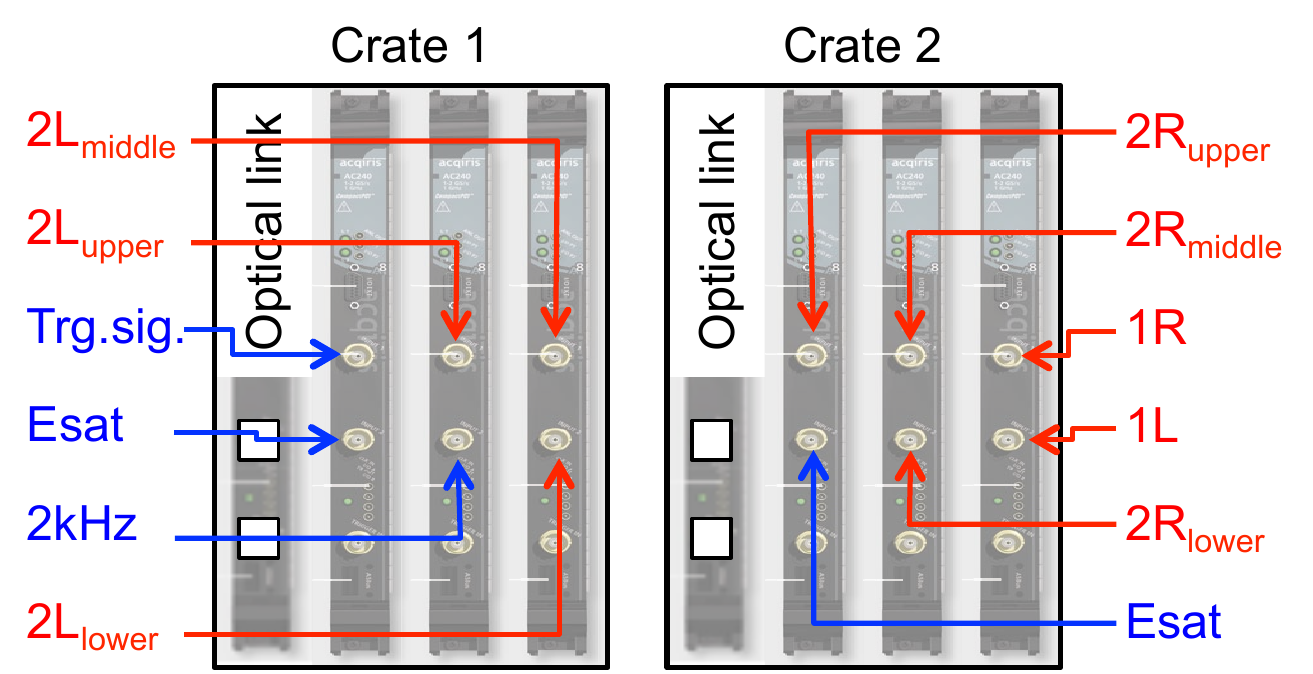}
\caption{\label{fig:pmtdaq} Schematics of the 2012 ICARUS PMT-DAQ specifically designed for neutrino time of flight
measurement.}
\end{figure}

The layout of the ICARUS T600 readout system is shown in Figure~\ref{fig:t600pmt}. A single cable is used to bias each PMT and extract the signal. The biasing circuit integrates a charge preamplifier, whose output is sent to the ICARUS trigger system. A fraction of the direct PMT signals ($\sim 1/10$) is extracted before the pre-amplification stage; the sum of these signals for each PMT horizontal array is sent to a DAQ system specifically designed for the neutrino time of flight measurement (Figure~\ref{fig:pmtdaq}).

This additional PMT-DAQ system, derived from the DAQ developed for the WArP experiment~\cite{[15]}, is based on six 2-channel, 8-bit, 1-GHz AGILENT AC240 digitizer boards~\cite{[08]}. The eight PMT-Sum signals are continuously read out and stored in a circular memory buffer of a depth of 8 kB/channel, sampled at 1 GHz. The full dynamic  range of the digitizers can be set as low as 50 mV, allowing to get a sensitivity of about 2 phe/ADC count. The digitizer boards are organized in two crates, connected to the event building computers and to data storage via optical link. All digitizers within the same crate are synchronized with each other at a 10 ps level.

At each CNGS trigger, the content of the circular buffer is frozen and additional data are read out for a total memory depth equivalent to 1.4 ms (slightly more than a full LAr-TPC drift time to include all possible scintillation signals produced along with the drifting ionization charge) and stored in a second local memory buffer. The content of all buffers is then transferred to the ICARUS data storage. Internal time stamping available in the digitizers is also stored allowing synchronization with the associated event imaging.

In the additional four channels, several TTL timing signals are recorded: the ESAT PPmS as in 2011 set-up, a 2 kHz synchronization pulse from the WR system in phase with the related PPS, a pulse generated in the ICARUS trigger box at trigger level. The latter is replicated through a dedicated fan-out setup and in order to be sent to the HPTF and to WR facilities for time stamping. The jitter between replicas have been certified to be of the order of 0.1 ns. All these signals allow defining an absolute time for the PMT-DAQ digitizers in the time bases related to the four different timing systems.  Moreover, the 2 kHz WR signal allows monitoring the stability of the digitizer frequency (at a level better than 1 ppm). A comparison of the WR time stamp and the 2 kHz signal allows also monitoring the stability of the WR protocol.

The synchronisation between the two PMT-DAQ crates is ensured by the ESAT PPmS signal which is send simultaneously to both crates. The jitter between the two copies of the ESAT PPmS signal has also been certified to be of the order of 0.1 ns.

\begin{figure}[htbp]
\centering 
\includegraphics[width=0.90\textwidth]{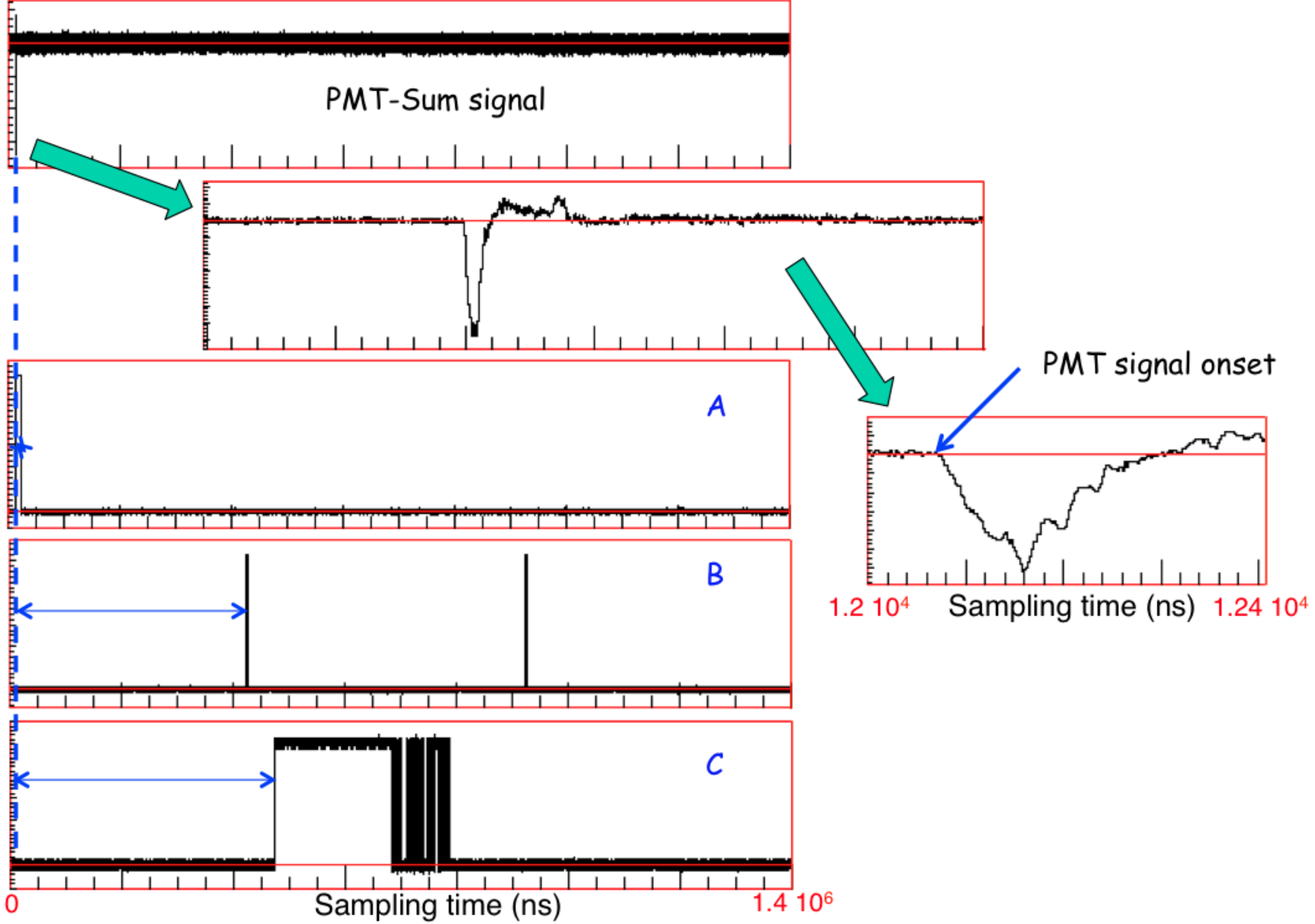}
\caption{\label{fig:pmtsig} Example of signals recorded on the ICARUS PMT-DAQ: (A) TTL signal generated in the
ICARUS trigger box (5 ns front edge, 10 $\mu$s wide): also sent to HPTF and WR-DAQ for time stamp; (B)
2 kHz signal from WR fine-delay (1 $\mu$s wide, 3 ns front edge): defines absolute time in WR time bases
and monitors Acqiris time stability; (C) 1 kHz (PPmS) ESAT timing signal (200 $\mu$s wide, 3 ns front edge,
followed by time encoding): defines absolute time in LNGS time base.}
\end{figure}

The position of the timing signals along the recorded waveforms is taken, by interpolation, at the signal leading edge with the same threshold as that used in the related timing systems for time stamping and in any calibration procedure. The time of the PMT-Sum is taken at the onset of the related signal, thus providing the arrival time of the earliest scintillation photons. An example of these signals is shown in Figure~\ref{fig:pmtsig}. 

\subsection{Calibration of the ICARUS PMT signal propagation}
In order to measure the neutrino arrival time in the ICARUS detector, the propagation time of the scintillation light signals from the PMTs to the AC240 boards, including the transit time within the PMTs, the overall cabling ($\sim 44$ m) and the delay through the signal adders, have to be calibrated.

The propagation along the cabling has been measured with an accuracy of  $\sim 0.5$ ns by means of standard reflection techniques on sharp signals (few ns rise-time) with a 1 GHz bandwidth oscilloscope. All 74 PMT cables resulted of the same length, equivalent to 233 ns within the precision of the measurement. 

The measurement of the PMT transit time was carried out in laboratory tests at room temperature on an ETL 9357FLA spare sample. The PMT was provided with the same voltage divider (base) adopted for the devices mounted in the East cryostat, where the Cathode to first dynode voltage difference is kept constant at +600 V. Two independent HV power supplies were used, one for the cathode and one for the dynode-chain and the PMT output was directly derived from the line to the anode by means of a 10 nF decoupling capacitor. In such way it was possible to disentangle the transit time contribution due to the windows focusing region from the same induced by the dynode-chain. This set-up allows also reproducing the voltage divider configuration of the West module, where a resistive drop on the voltage divider base defines the cathode to first dynode voltage difference.

As a light source, a laser diode was used, excited using a fast pulser (1 ns rise time, 4 ns width). The light was brought to the photocathode facing directly the laser diode to the PMT window.
For each event, the pulser signal was used as a trigger of a 1 GHz oscilloscope. For each pulse the transit time was evaluated from the delay of the onset of the PMT signal, corresponding to the arrival time of the earliest prompt photons, with the same method used for the T600 data. Data were corrected by the delay induced by the cables and the laser diode-PMT distance.

\begin{figure}[htbp]
\centering 
\includegraphics[width=0.45\textwidth]{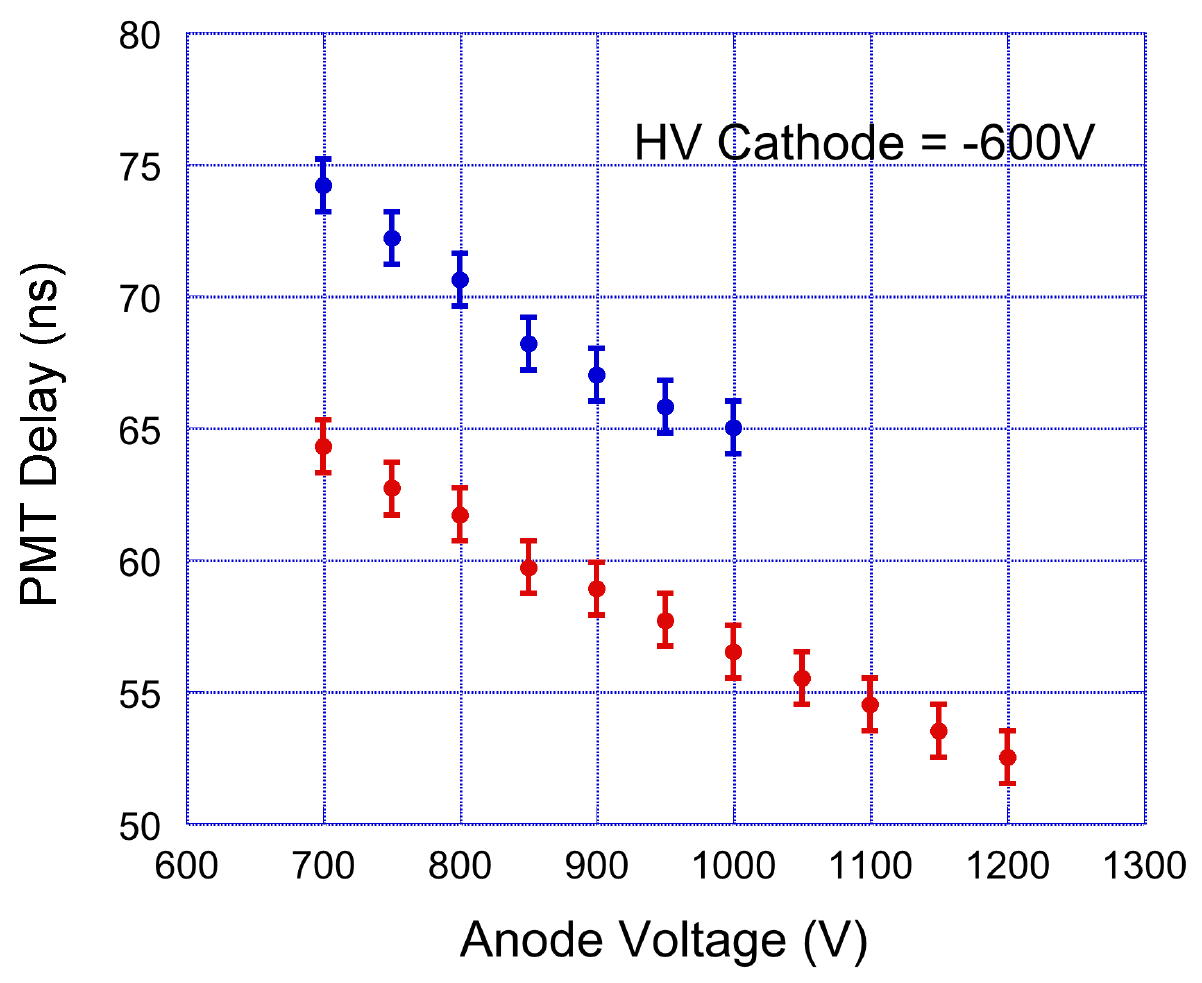}
\includegraphics[width=0.45\textwidth]{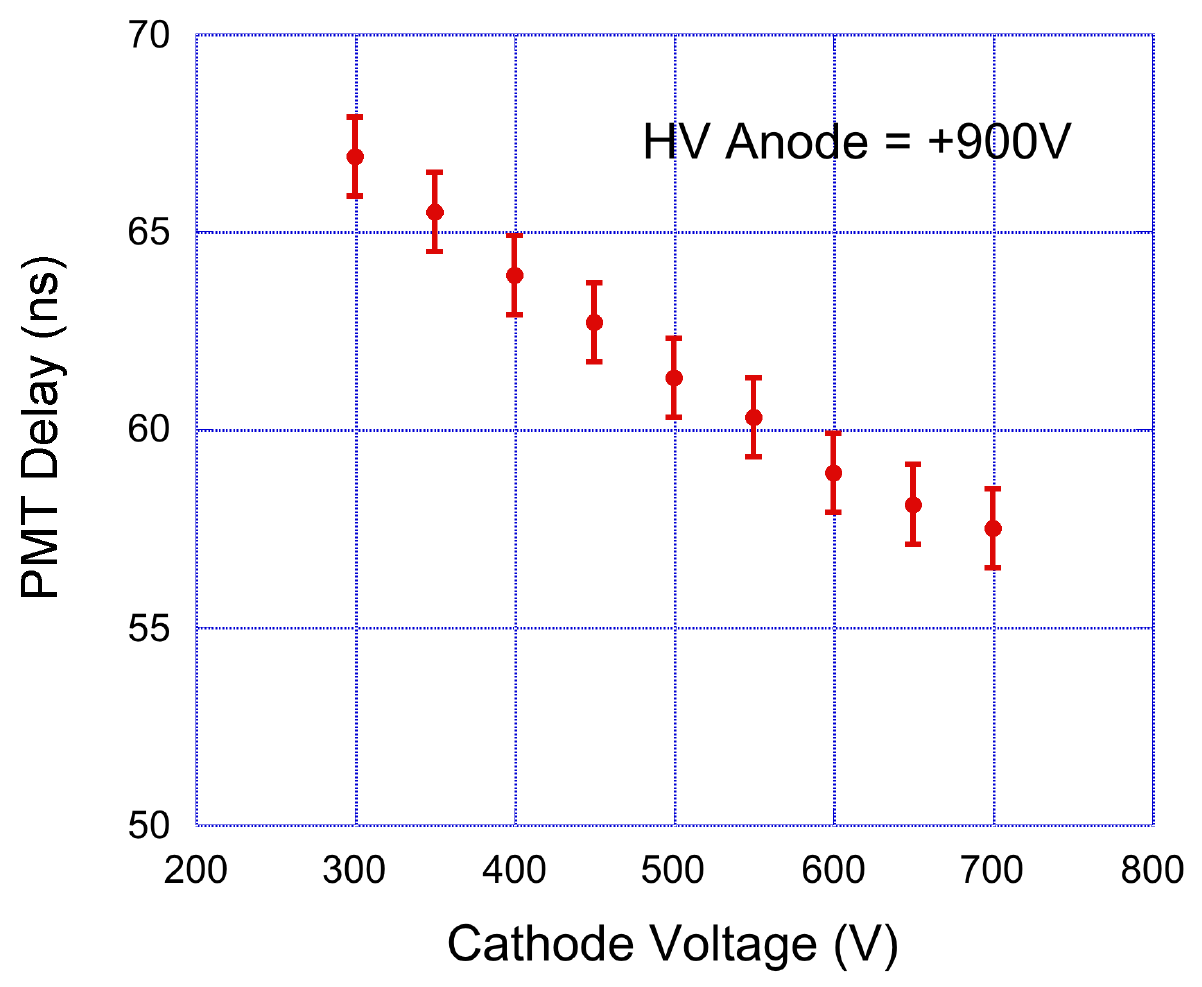}
\caption{\label{fig:pmtcal} PMT transit time measurements in laboratory tests. Red: PMT installed horizontally along the
East-West direction, as in the T600 cryostats; blue: PMT installed vertically to enhance the effect due to
earth magnetic field. Left: configuration with fixed cathode to first dynode voltage difference (East module
configuration). Right: configuration with fixed Anode to first dynode voltage difference. The combination
of left and right plots allows calculating the transit time for the West module voltage divider configuration.}
\end{figure}

In order to minimize variations on the transit time due to the earth magnetic field, the PMT under test was positioned horizontally along the East-West direction, as in the T600 cryostats. In this condition the PMT was exhibiting the highest gains and the minimum transit time values (Figure~\ref{fig:pmtcal}).

In dedicated tests at cryogenic temperature (~\cite{[16]}) only a reduction of the transit time jitter was observed at 77 K (1.3 ns) with respect to room temperature (2.5 ns).

\begin{figure}[htbp]
\centering 
\includegraphics[width=0.90\textwidth]{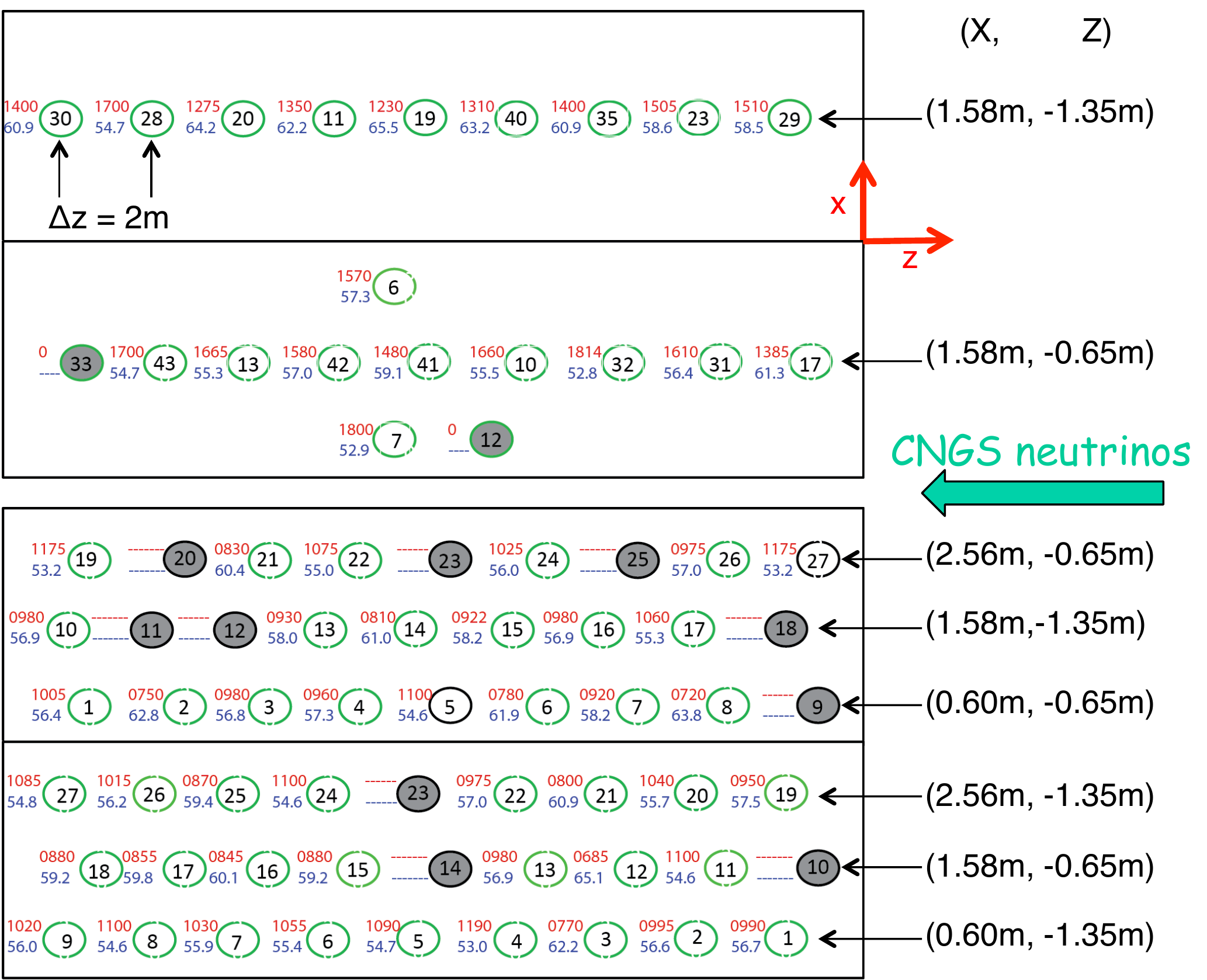}
\caption{\label{fig:pmtmap} Layout of the PMT's in the ICARUS T600 detector. Biasing voltage (V, red) and transit time
(ns, blue) are also shown. Position (m) of the most upstream PMT's with respect to the CNGS beam is
given in the ICARUS coordinate system (X-Z plane). PMT out of service are marked in dark grey.}
\end{figure}

Being the PMT transit time a purely geometrical and electrostatic effect, the laboratory measurement allowed determining the transit time of all the PMT installed in the T600, given the applied biasing voltage. The residual associated error of 1 ns is related to the oscilloscope sampling frequency of 1 GHz. Figure~\ref{fig:pmtmap} shows the PMT map in the ICARUS T600 detector, including biasing voltage and transit time.

\subsection{Calibration of timing signals}
For the classic 2011 timing system and the HPTF facility, the propagation delays of the signals along the $\sim 8$ km optical fibres connecting the external laboratory and the underground halls have to be determined. The method consists in measuring the time difference $\Delta t$ and time sum $\Sigma t$ of the signal propagation along the usual path and an alternative one consisting a spare fibre. For the ESAT signal:
\begin{itemize}
\item in the first configuration the 1PPS output of the ESAT-2000 GPS receiver was converted into an optical signal, sent underground via the spare fibre and converted back into electrical. The difference in the propagation delays, between this signal and the 1PPmS output of the ESAT-2000 GPS propagated over the standard path, was measured underground taking as a reference the middle height of the rising edge;
\item in the second configuration the 1PPmS output of the ESAT-2000 GPS, at the end of the usual path underground is sent back to the external laboratory, where it is compared with the 1 PPS signal, taking as a reference the middle height of the rising edge.
\end{itemize}

The used optoelectronic chain is kept identical in the two cases by simply swapping the receiver and the transmitter between the two locations. Furthermore the jitter of the phase difference between the 1PPS and the 1PPmS outputs of the ESAT-2000 GPS receiver was checked to be negligible ($\lesssim 0.25$ ns).

This procedure was applied several times before and after the 2012 bunched beam run, obtaining a result of $41905.3 \pm 2.0$ ns where the error takes into account  the jitter in the various measurements and the observed response variation of the receivers electronics and of the signal propagation time along the fibre mainly due to temperature changes.  An equivalent method has been used for the optical fibre transferring the ICARUS trigger related pulse to the HPTF ($45260.8 \pm 1.0$ ns); the smaller associated error is mainly due to the improved receivers electronics. 

As a reminder, the WR timing system does not need this calibration, which is intrinsically and continually performed by the WR protocol. On the other hand, and for the sake of reducing the number of elements in the delay path, which would include PolarRx2 PPS output to WR reference node, WR reference node to WR acquisition node and WR acquisition node to acquisition scope, it is convenient to perform a direct two-way calibration between the acquisition digitiser and the PolarRx2 PPS output. Notice that as the PolarRx2 receiver and the WR system are clock synchronous with the CS4000, and as the WR system compensates for variations in fiber delay, the PolarRX2 PPS output to WR Scope timestamp delay drift will mainly depend on the electronics temperature changes and voltage changes. Laboratory tests in a thermalized chamber have shown that variations in the WR system are smaller than 300 ps peak-peak for the -20$^\circ$C to 40$^\circ$C range~\cite{[11]}.

\section{Geodetic measurement of the CERN-LNGS distance}
\label{sec:geodesy}
Fundamental for the neutrino velocity measurement is the knowledge of the distance between the point where the proton time-structure is measured at CERN and the reference point at the entrance of the ICARUS detector in the LNGS underground halls.

The distance between the CNGS target focal point and the LNGS experiment has been remeasured under the responsibility of the Politecnico di Milano just before the 2012 bunched beam run, following a dedicated geodesy campaign. The 3D coordinates of several reference points of various LNGS experiments were measured by establishing GPS benchmarks at the two sides of the $\sim10$ km long Gran Sasso highway tunnel and by transporting their positions with a terrestrial traverse down to the underground detectors sites, with an estimated accuracy of 3.7 cm. An equivalent accuracy is provided for CNGS target focal point in the common geodetic reference frame. 

\begin{figure}[htbp]
\centering 
\includegraphics[width=1.00\textwidth]{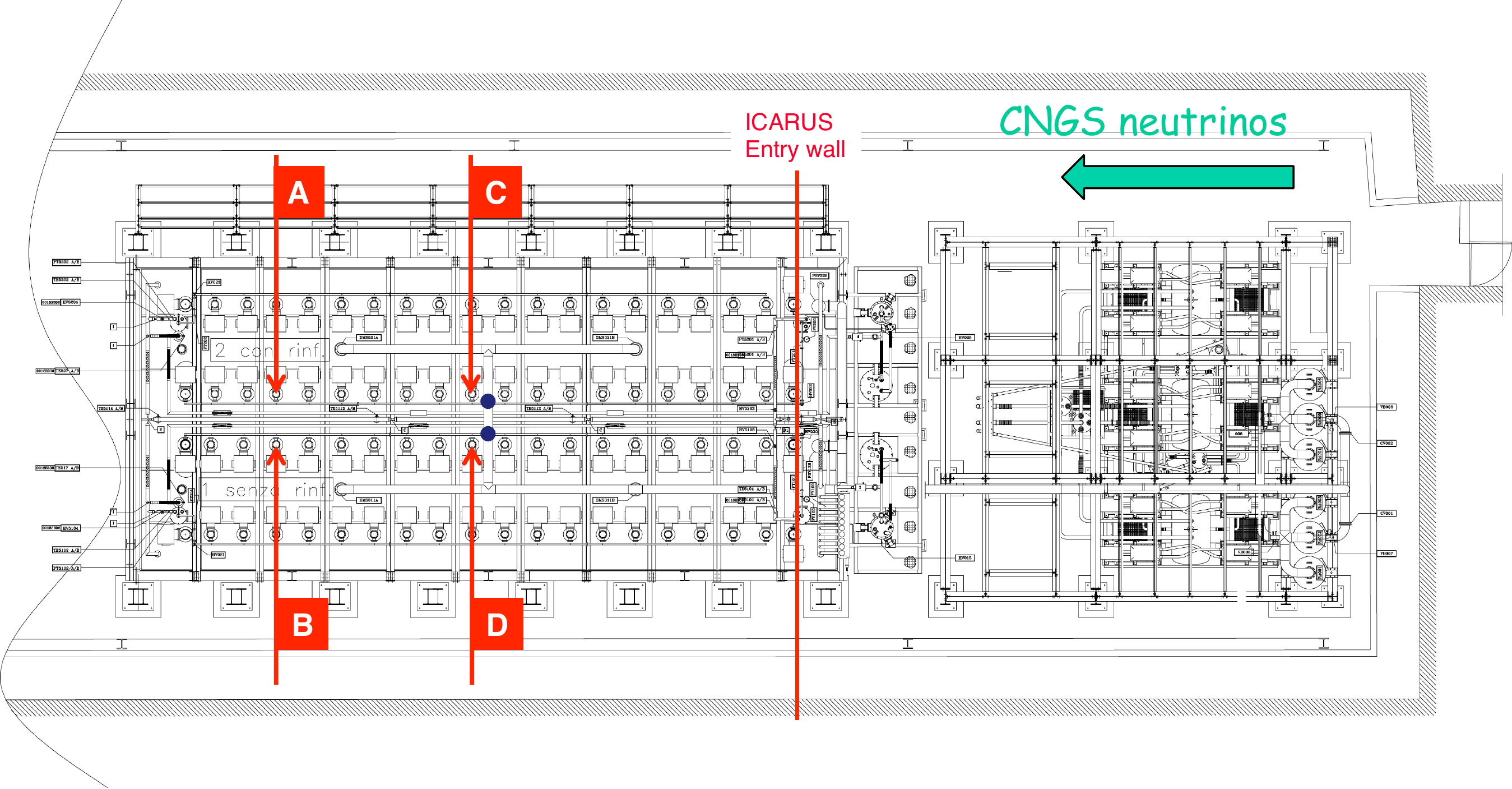}
\caption{\label{fig:t600top} Top view of the ICARUS detector in the LNGS underground Hall B. The four points, (A, B, C and D) have been measured in the Politecnico di Milano geodetic campaign. The solid blue circles indicate the position on the floor of the two detector ``fixed point''. The ICARUS entry wall used as ``reference entry point'', as defined in the text, is also shown.}
\end{figure}

In the case of the ICARUS T600 detector four points were measured on the top floor of the detector, where the racks of the front-end electronic are located, close to the two ``fixed points'' of the detector (see Figure~\ref{fig:t600top}).

The aluminium cold body of each module of the ICARUS T600 detector is fixed to the floor of the Hall B on a single ``fixed point'', located half a way of the longest side and in correspondence of the innermost wire chamber. The other feet, over which the detector is positioned, allow displacements along the horizontal plane in view of the several cm contractions during cooling to LAr temperature. The stainless steel inner detector structure is also fixed to the cold body only at the same ``fixed point''  to take into account the different contractions of stainless steel and aluminium. As a consequence, from the four reference points on top of ICARUS the coordinated of the two fixed points have been evaluated and, from these, the 3D position of the CNGS neutrino entry point in the LAr active volume on the inner detector structure has been derived. To take into account also the 52 mrad inclination of the neutrino beam with respect to the Hall B floor, the ICARUS ``reference entry point'' has been chosen at half height of the active volume.

In this framework the distance from the CNGS target focal point to the ICARUS ``reference entry point'' is $730478.56 \pm 0.09$ m where the error includes the uncertainty in the propagation from the measured points to the ``reference entry point'' (including contraction of cold body and inner detector).

Adding the precisely known distance from BCT position and CNGS target focal point $743.391 \pm 0.002$ m, the baseline considered for the measurement of the neutrino velocity is then $731221.95 \pm 0.09$ m, in agreement with the 2011 estimation $731222.3 \pm 0.5$ m extrapolated from data of an older geodetic campaign performed for the OPERA experiment.
As a consequence, the expected time of flight for v = c is $2439096.08 \pm 0.3$ ns, including the 2.2 ns contribution due to earth rotation (Sagnac effect). 

\section{Data analysis}
\label{sec:analysis}
During the 2012 two weeks of data taking with the CNGS in bunched mode, the ICARUS T600 detector collected 25 beam-associated events, consistent with the CNGS delivered neutrino flux of $1.8 \times 10^{17}$ p.o.t. 

\begin{figure}[htbp]
\centering 
\includegraphics[width=1.00\textwidth]{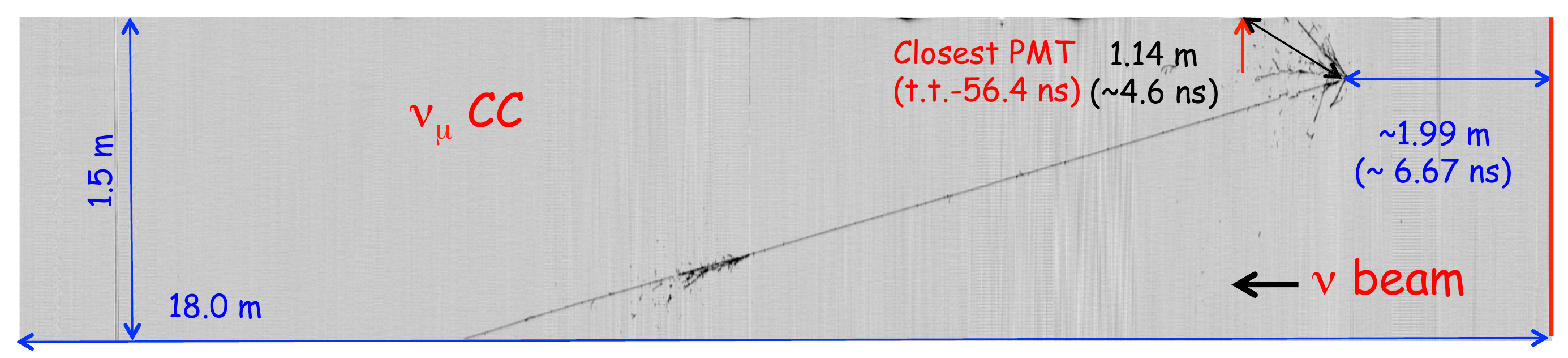}
\includegraphics[width=1.00\textwidth]{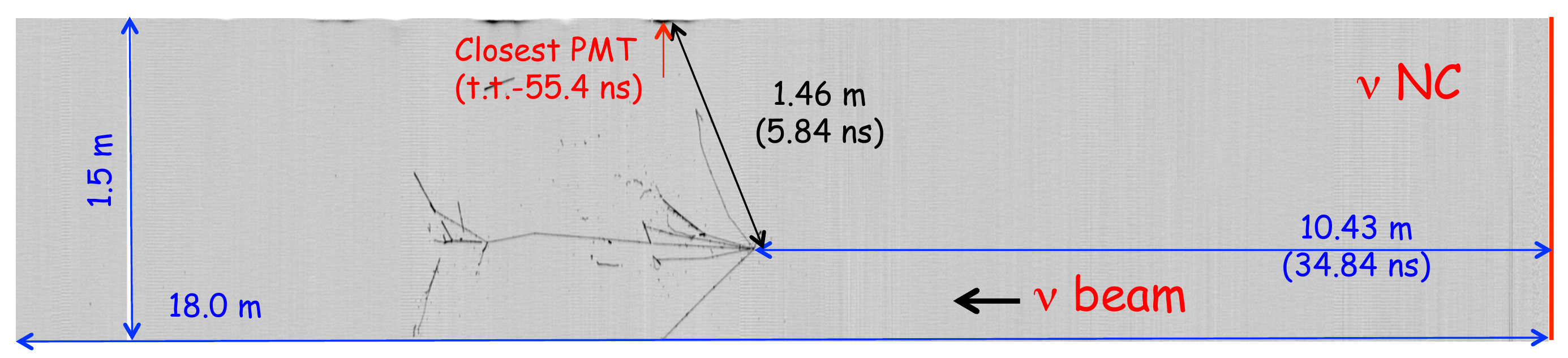}
\includegraphics[width=1.00\textwidth]{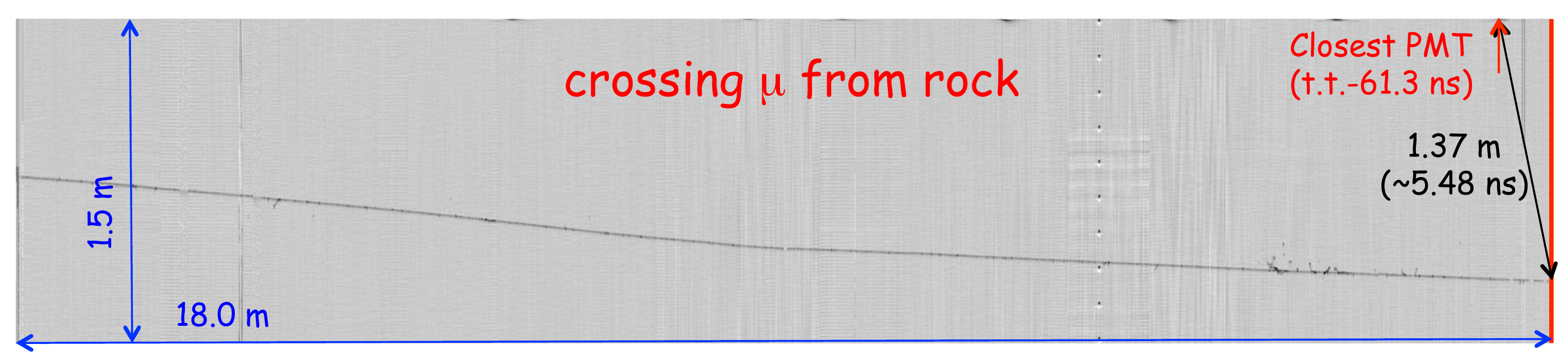}
\includegraphics[width=1.00\textwidth]{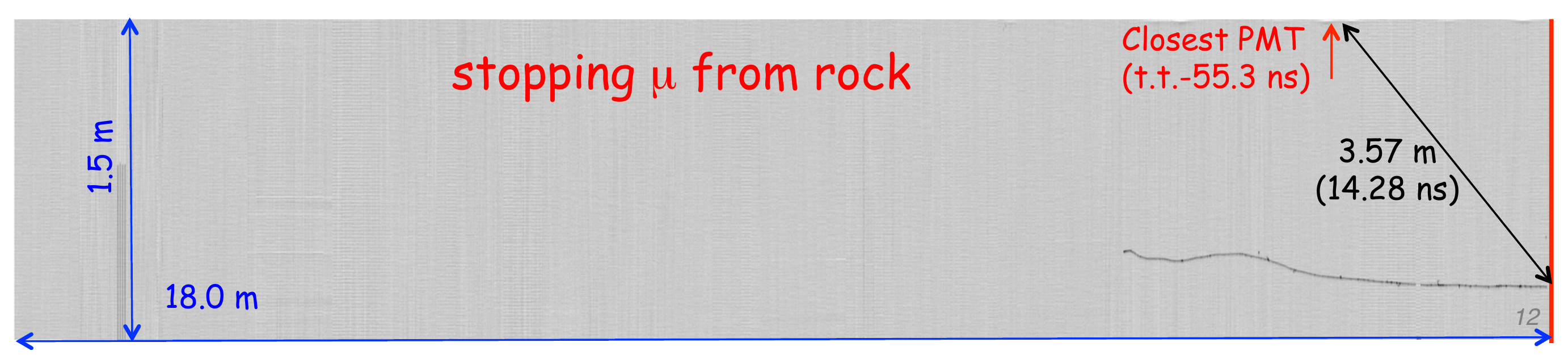}
\caption{\label{fig:events} Examples of different event topologies in the ICARUS LAr-TPC from the 2012 CNGS
bunched beam run. A 2D view (``Collection'' wire plane) is shown. For each event the distance from the ``ICARUS reference
entry point'' (red line on the right of the image) is reported as well as the photon path the nearest PMT
and the associated transit time (t.t.) in the PMT itself.}
\end{figure}

The events consist of 8 neutrino interactions (six CC and two NC) with vertex contained within the ICARUS fiducial volume and 17 additional through going muons (one of which stops within the LAr active volume) generated by CNGS neutrino interactions in the upstream rock.  Events in the standard ICARUS DAQ and the PMT-DAQ have been associated through their common absolute time-stamp. In Figure~\ref{fig:events} a sub-sample of these events is shown. 

For all the collected events, the visual scanning and 3D reconstruction are exploited to determine the distance of the interaction vertex from the ICARUS ``reference entry point'' and to quantify the shortest path of the scintillation light to the nearest PMT. In LAr the 128 nm scintillation photons propagate at 4.0 ns/m (refraction index = 1.20) with a Rayleigh scattering length as large as 80 cm hence comparable with the typical distance from ionization track to the PMTs. Given the ``millimeter'' space resolution of the ICARUS LAr-TPC, both the above corrections are measured with sub-ns accuracy. In the examples of Figure~\ref{fig:events}, these corrections are also reported.

With reference to the layout in Figure~\ref{fig:layout} and~\ref{fig:t600pmt}, the actual neutrino time of flight is measured on an event-by-event basis: (1) the transit time of the proton beam bunch at the BCT in the CERN time-base is used as starting time; (2) the neutrino arrival time is taken at the ICARUS ``reference entry point'' in the LNGS time base; (3) the alignment of the LNGS and CERN time bases is calculated following the different synchronization paths available for the 2012 run. The first term is determined picking up in the BCT waveforms the proton bunch giving the neutrino time of flight value closest to expectation for v = c.

\begin{figure}[htbp]
\centering 
\includegraphics[width=1.00\textwidth]{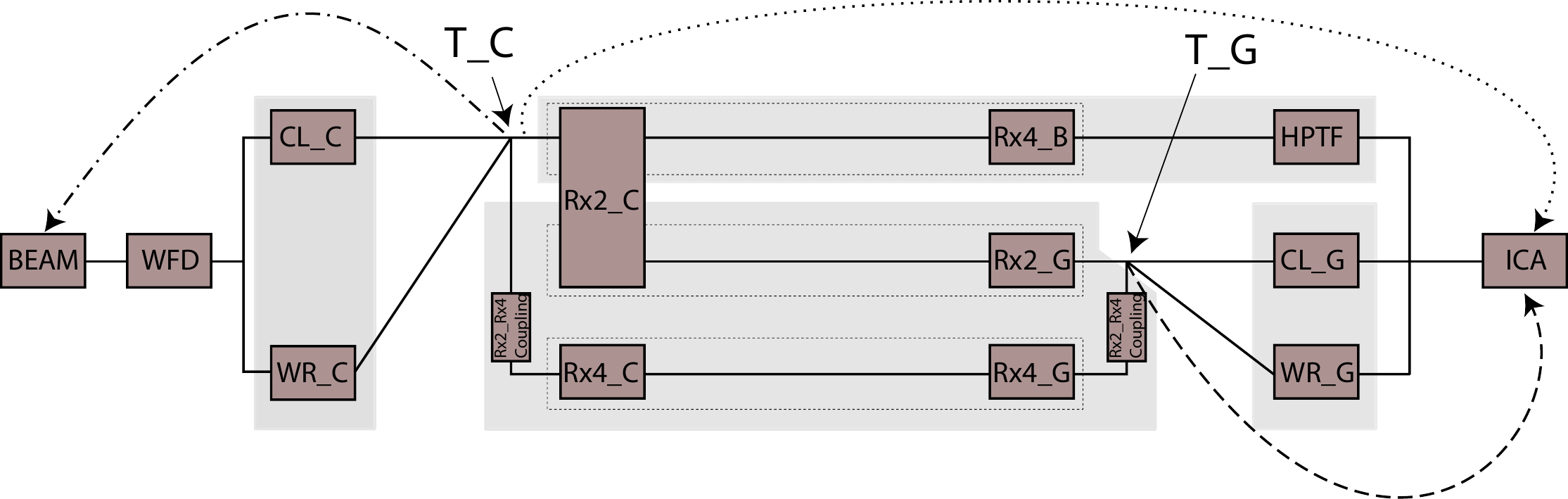}
\caption{\label{fig:delaypaths_new} Schematics of the averaging procedure of the synchronization paths, used to improve the accuracy of the neutrino time of flight measurement. Legenda: $BEAM$: BCT proton timing; $WDF$: Proton beam waveform digitiser; $CL\_C$: CERN classical path with CTRI's; $WR\_C$: CERN White Rabbit path; $RX4\_B$: PolarRx4 in Borexino HPTF; $CL\_G$: LNGS classical path CTRI + LNGS timing; $WR\_G$: LNGS White Rabbit path; $HP\_B$: HPTF Borexino timing; $ICA$: Icarus detector response; $T\_C$: time stamp of the proton transit time at CERN; $T\_G$: time stamp of the neutrino arrival at LNGS from Classic and WR paths (for HPTF path the time stamp of neutrino arrival is referred directly to $T\_C$). Shaded areas indicate independent averaging stages.}
\end{figure}

With respect to the 2011 campaign, the accuracy of the experiment  has been improved by combining, also on an event-by-event basis, the timing measurements performed at each stage of the above described procedure, as depicted in the schematic diagram presented in Figure~\ref{fig:delaypaths_new} and according to the prescriptions described in  Ref.~\cite{[19]}. The time stamps of the proton transit time (recorded in the position marked as a $T\_C$) and that of the neutrino event arrival in ICARUS  ($T\_G$ in the picture) are the weighted averages of the measurements performed through the classic and WR paths. Also the link between CERN and LNGS time-bases has ben determined as weighted average of the time-transfer corrections available through the CERN-PolarRx2 to LNGS-PolarRx2  and CERN-PolarRx4 to LNGS-PolarRx4 GPS receivers paths, which were cross-calibrated by the METAS Institute~\cite{[17]}  and independently checked by the PTB Institute~\cite{[18]} . The HPTF branch provides timestamps which  already include the CERN-LNGS time-transfer through the CERN-PolarRx2 to HPTF-PolarRx4e GPS receivers path, which was cross-calibrated by INRIM and ROA Institutes~\cite{[12]}. 

The main calibration parameters and the associated sources of systematic and statistical errors, discussed is the previous sections, are summarized in Table~\ref{tab:1}.  The systematic errors of each branch of the measurement have been used to perform the averages along the different timing paths. Cross-correlations, essentially due to the common use of the same GPS signals by the various receivers and the use of the same PolarRx2 receiver at CERN by HPTF and the Classical/WR path at LNGS, have also been considered.
The final timing is obtained as weighted average of the HPTF timing with that obtained with the alternative paths.
The overall systematic error derived within the averaging procedure is $\sim 2.39$ ns. The expected event distribution width is  $\sim 3.0$ ns, dominated by the detector response (mainly related to PMTs), the width of the proton beam bunches and residual synchronization jitter.

\begin{table}[tbp]
\centering
\begin{tabular}{|l| c c c|l|c|}
\hline
 {\scriptsize BCT delay (ns) }	&						& {\scriptsize 583.68$\pm$1.0}&                              &{\scriptsize Syst.}   &{\scriptsize Dedicated meas. ($+$Cs) }  \\
 {\scriptsize Beam pulse width (ns)} &				& {\scriptsize  $\pm$1.8}&                              &{\scriptsize Stat.}   &{\scriptsize Digitizer measurement}  \\
 \hline
{\scriptsize CERN UTC distrib. } 	&   {\scriptsize   \sl Classic  }	&		&{\scriptsize \sl WR }             &                      &{\scriptsize Portable  Cs $+$ }   \\
{\scriptsize + WFD trigger (ns)}	&  {\scriptsize  10106.1$\pm$2.0 }&                               &{\scriptsize $\pm$1.0 }      &{\scriptsize Syst.}   &{\scriptsize Two-ways $+$ Scope  }\\
 \hline
{\scriptsize CERN-LNGS timing    }	&  {\scriptsize   \sl PolarRx2   }    &{\scriptsize \sl PolarRx4  }       &{\scriptsize \sl HPTF    }        &                      &{\scriptsize METAS$+$PTB and}   \\
{\scriptsize  intercalibration$^a$ (ns)}	&  {\scriptsize   2.3$\pm$2.0  }&{\scriptsize $\pm$2.1 }       &{\scriptsize $\pm$1.1$^b$ }      &{\scriptsize Syst.}   &{\scriptsize INRIM$+$ROA (HPTF)} \\
\hline 
{\scriptsize LNGS UTC distrib.} 	&   {\scriptsize \sl Classic}         &  {\scriptsize  \sl WR}            &{\scriptsize  \sl HPTF}          &                      &                   \\
{\scriptsize on optical fiber (ns)}	&  {\scriptsize 41905.3$\pm$2.0}&  {\scriptsize $\pm$1.0}     &{\scriptsize  45260.8$\pm$1.0}&{\scriptsize Syst.}   &  {\scriptsize Two-ways }                 \\
\hline
{\scriptsize  Overall residual } 	&                                 &                                                 &                              &                                    &   {\scriptsize Comparison of individual}  \\
{\scriptsize synchronization jitter (ns)}	&                                 &  {\scriptsize $\sim$1.0}        &                              &   {\scriptsize Stat.}   &   {\scriptsize timing paths}      \\
\hline
{\scriptsize ICARUS PMT's: }	&                                 &                               &                              &                      &                  \\
{\scriptsize Transit time (ns)}	&                                 &{\scriptsize $\pm$1.0}         &                              &{\scriptsize Syst.}   &{\scriptsize Dedicated measurement}  \\
{\scriptsize Cable lenght (ns)}	&                                 &  {\scriptsize 233.0$\pm$0.5}    &                              &{\scriptsize  Syst.}  &{\scriptsize Signal reflection (Scope)} \\ 
{\scriptsize Pulse identif. and}   &  	&               &                              &  &{\scriptsize Visual scanning $+$} \\
{\scriptsize topology corr. (ns)} &                              &         {\scriptsize $\pm$2.1}     &                              &      {\scriptsize  Stat.}     &{\scriptsize PMT time jitter}   \\
\hline
{\scriptsize CERN-BCT to T600}     &                             &                                                             &                              &                                  &                                    \\
{\scriptsize ref. point distance (ns)}&                             & {\scriptsize 2439096.1$\pm$0.3}  &                              &  {\scriptsize Syst.}  &  {\scriptsize Geodesy}   \\
\hline     
\end{tabular}
\caption{\label{tab:1} Summary of the main calibration parameters and the associated sources of statistical and systematic errors. Notes: (a) the METAS+PTB and INRIM+ROA CERN-LNGS inter-calibration systematics include cross correlation terms, estimated to be of the order of  $0.5 - 1.0$ ns, due to the use of the same GPS signals; (b) the used calibration strategy absorbs the fixed delays in the time dependent atomic clock alignment values.}
\end{table}

\begin{figure}[htbp]
\centering 
\includegraphics[width=0.60\textwidth]{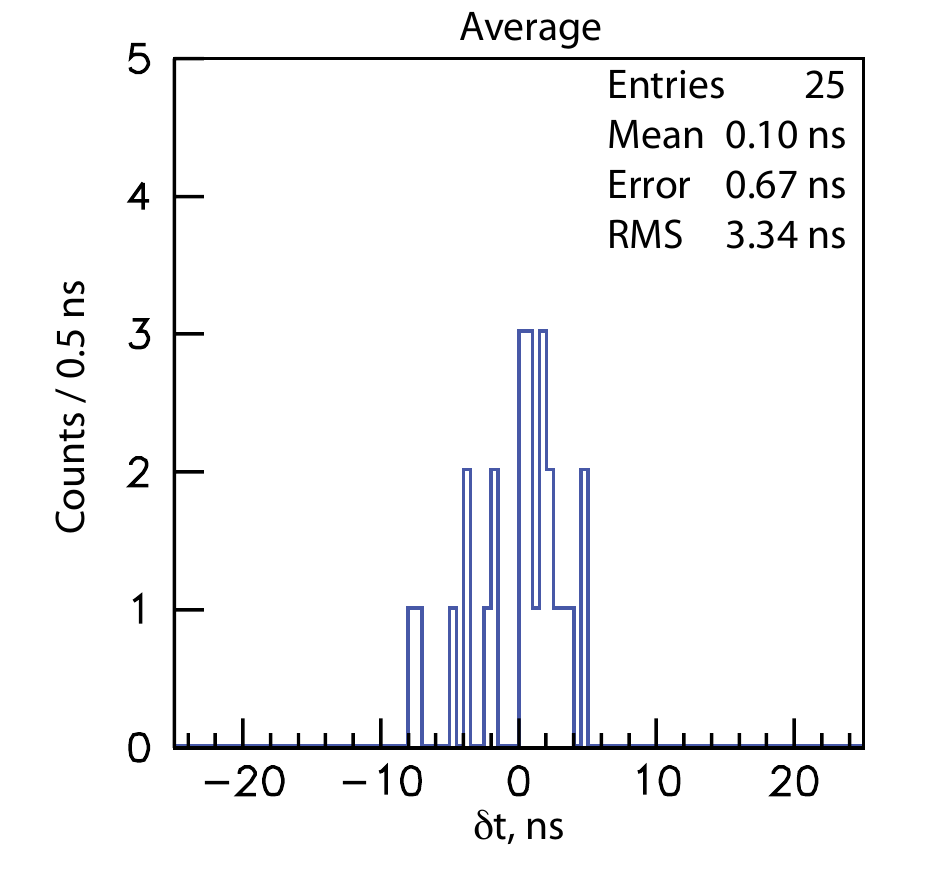}
\caption{\label{fig:tof_aver} Event distribution in ICARUS T600 for $\delta$t = tof$_c$ - tof$_\nu$, according to the averaging procedure of all synchronisation paths described in the text.}
\end{figure}

The difference between neutrino time of flight (tof$_\nu$) from the BCT to the ICARUS ``reference entry point'' and the expected time of flight based on the speed of light (tof$_c$ = 2439096.1 ns) is shown in Figure~\ref{fig:tof_aver}. The resulting value $\delta t = tof_c - tof_\nu = 0.10\pm0.67_{stat.}\pm2.39_{syst.}$  ns, is fully compatible with the neutrino propagation at the speed of light. The corresponding deviation of the neutrino velocity from the speed of light is $\delta (v/c) = (v_\nu - c)/c = (0.4\pm2.8_{stat.}\pm9.8_{syst.}) \times 10^{-7}$, excluding neutrino velocities larger than the speed of light by more than $1.35\times10^{-6} c$ at 90$\%$ C.L..

\begin{figure}[htbp]
\centering 
\includegraphics[width=0.40\textwidth]{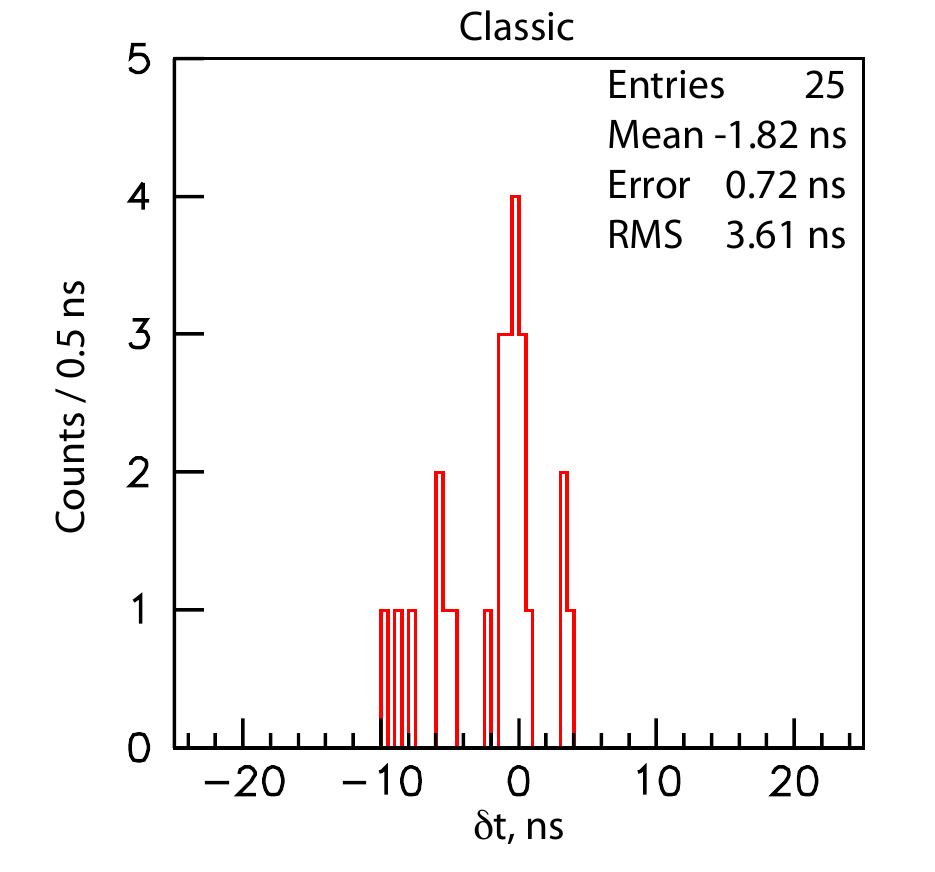}
\includegraphics[width=0.40\textwidth]{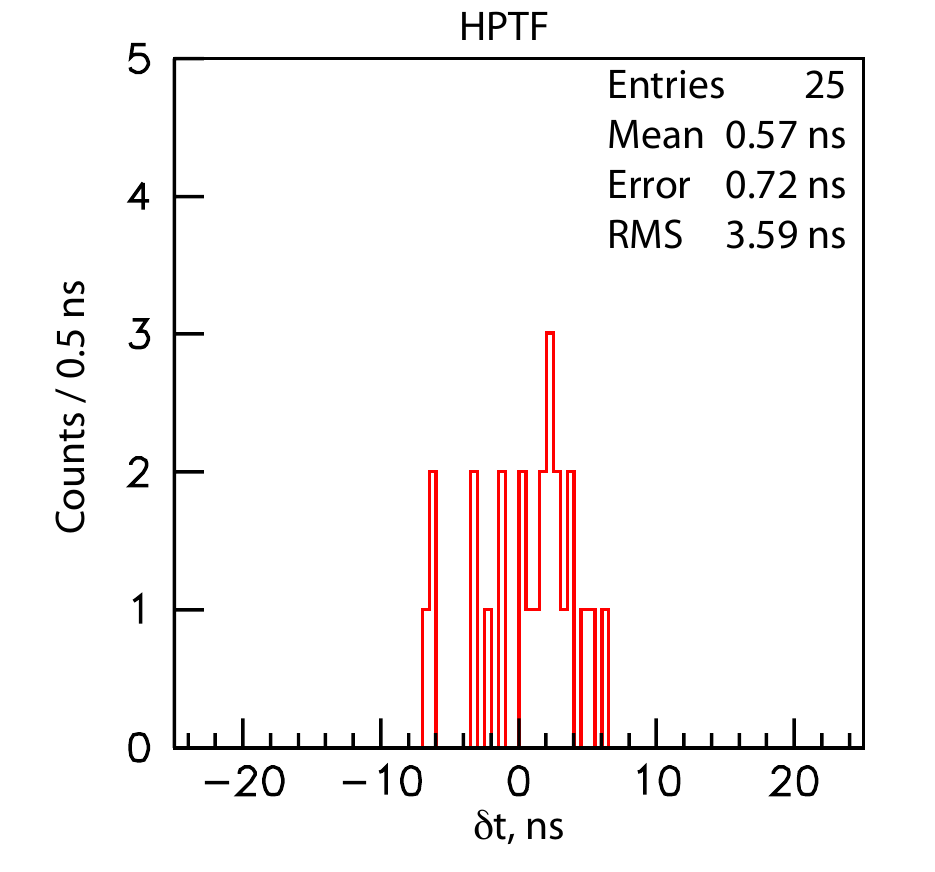}
\includegraphics[width=0.40\textwidth]{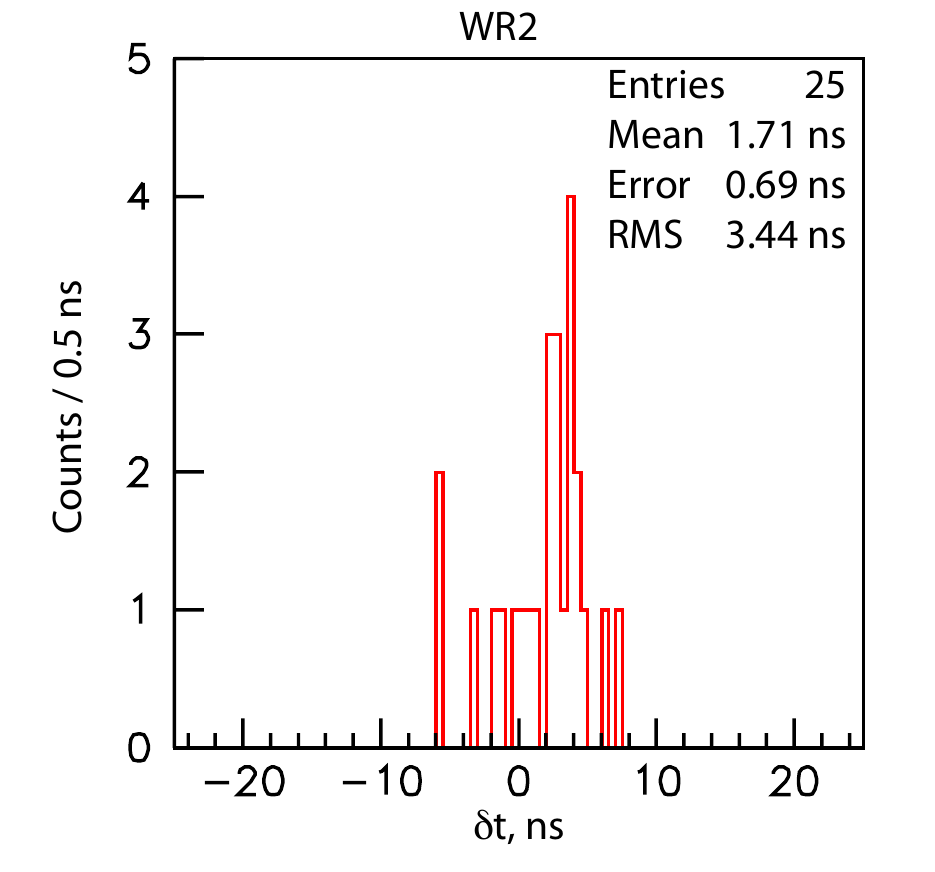}
\includegraphics[width=0.40\textwidth]{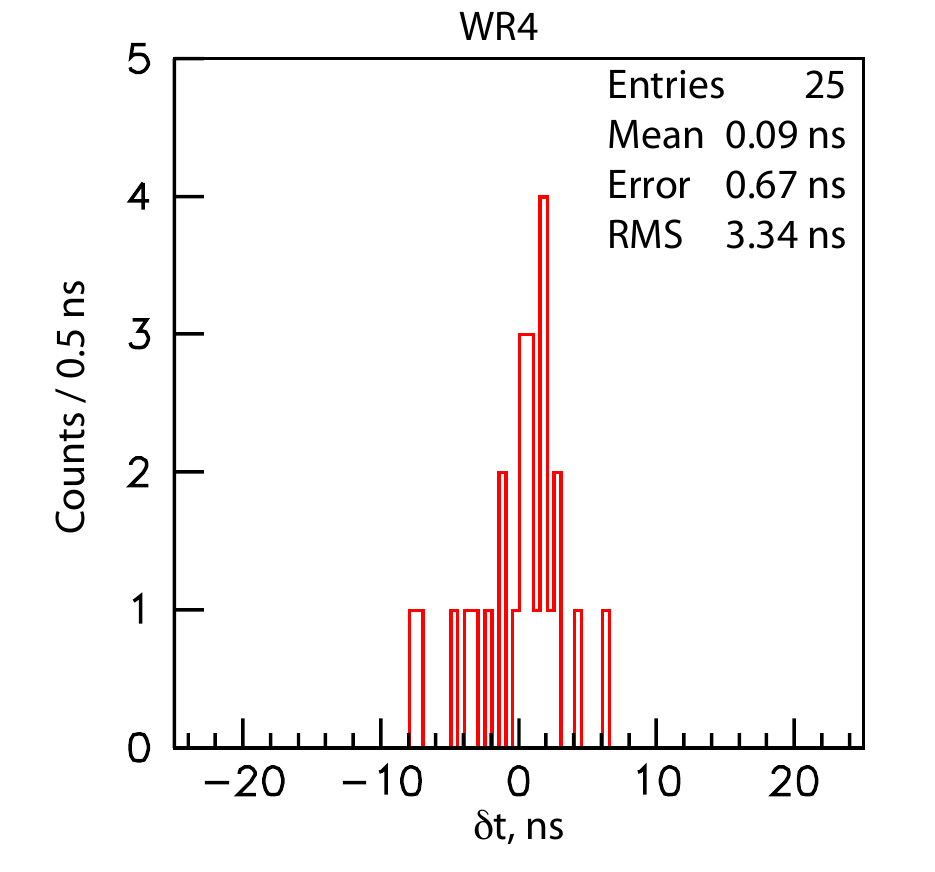}
\caption{\label{fig:tofc-nu} Event distribution in ICARUS T600 for $\delta$t = tof$_c$ - tof$_\nu$ according to the four different synchronization
paths. Classic: 2011 path; HTPF: Borexino timing facility at LNGS and Classic layout at CERN;
WR-2: White-Rabbit setup through PolarRx2 GPS receivers; WR-4: White-Rabbit setup through PolarRx4 GPS receivers.}
\end{figure}

Only for a more direct comparison with previous measurements and with the results of the other experiments, the distributions of the neutrino time of flight difference is shown in Figure~\ref{fig:tofc-nu}, for four specific synchronization paths. As expected the distribution widths are slightly larger that the one of the average and very similar for all paths, indicating that indeed the jitter is dominated by detector response and the width of the proton beam and the residual synchronization jitter.  Morever the mean values of the four paths are consistent to each other and with that of the average distributions within  the quoted systematic error.

\section{Conclusions}
\label{sec:conclude}
ICARUS detector was exposed in May 2012 to the especially designed CNGS neutrino bunched beam in order to measure with a high accuracy the CERN to LNGS neutrino time of flight. The detector recorded 25 neutrino beam associated events. The expected time of flight difference between the speed of light from CERN to ICARUS and the actual position of the vertex of the LAr-TPC events has been neatly analysed, leading to the result $\delta t  = tof_c - tof_\nu = 0.10\pm0.67_{stat.}\pm2.39_{syst.}$ ns,  fully compatible with the neutrino propagation at the speed of light and in agreement with the 2011 ICARUS result, with improved statistical and systematic errors. This measurement excludes neutrino velocities exceeding the speed of light by more than $1.35\times10^{-6} c$ at 90$\%$ C.L.


\acknowledgments

The ICARUS Collaboration acknowledges the fundamental support of INFN and, in particular, of the LNGS Laboratory and its Director, to the construction and operation of the experiment. Moreover the authors thank the Re-search and Technical Divisions and the Computing and Network Service of LNGS for their contribution in preparing the timing facility; in particular, the cooperation of S. Parlati, G. Di Carlo and N. Taborgna is warmly recognized as well as that of A. Rubini (University of Pavia) for his help in debugging the WR system. The contribution of A. Razeto for preparing the HPTF facility is especially acknowledged. The authors thank LEICA Geosystems Italia for having supplied the instruments used in the geodetic campaign at LNGS and the LNGS technical staff for the logistic support. Moreover the contribution of the INRIM, ROA, METAS and PTB Institutions in the calibration of the timing facilities is recognized. The Polish groups acknowledge the support of the Ministry of Science and Higher Education in Poland, (project 637/MOB/2011/0) and National Science Center, Harmonia funding scheme. Last but not least, the contribution of CERN is gratefully acknowledged, in particular E. Gschwendtner with all the CNGS staff for the very successful operation of the neutrino beam facility and the Survey Team for the information on the target coordinates at CERN.



\end{document}